\documentclass[fleqn,usenatbib]{mnras}

\usepackage{newtxtext,newtxmath}
\usepackage[T1]{fontenc}

\DeclareRobustCommand{\VAN}[3]{#2}
\let\VANthebibliography\thebibliography
\def\thebibliography{\DeclareRobustCommand{\VAN}[3]{##3}\VANthebibliography}

\usepackage{graphicx}	% Including figure files
\usepackage{amsmath}	% Advanced maths commands
\usepackage{orcidlink}
\usepackage{anyfontsize}

\usepackage{soul}
\usepackage{ragged2e}
\definecolor{green2}{rgb}{0.15,0.6,0.05}
\definecolor{rust}{rgb}{0.7,0.1,0.1}

\usepackage{adjustbox}
\usepackage[super]{nth}
\usepackage{hyperref}
\usepackage{longtable}
\usepackage{booktabs}
\usepackage[percent]{overpic}
\usepackage{stfloats} % for positioning of figure* on the same page
\usepackage[caption=false]{subfig} 

\title[AT~2016blu Spectroscopy]
{Spectroscopy of AT 2016blu's recurring supernova impostor outbursts}

\author[M. Aghakhanloo et al.]{Mojgan Aghakhanloo \orcidlink{0000-0001-8341-3940},$^{1}$\thanks{E-mail:
mvy4at@virginia.edu} Nathan Smith \orcidlink{0000-0001-5510-2424},$^2$ Jennifer E. Andrews \orcidlink{0000-0003-0123-0062},$^{3}$ 
Alexei V. Filippenko \orcidlink{0000-0003-3460-0103},$^{4}$ \newauthor
Griffin Hosseinzadeh \orcidlink{0000-0002-0832-2974},$^{5}$
Jacob E. Jencson \orcidlink{0000-0001-5754-4007},$^{6}$ 
Jeniveve Pearson \orcidlink{0000-0002-0744-0047},$^{2}$ 
David J. Sand \orcidlink{0000-0003-4102-380X}$^{2}$ \newauthor 
Thomas G. Brink \orcidlink{0000-0001-5955-2502},$^{4}$
Kelsey I. Clubb $^{4}$ 
\\
$^1$ Department of Astronomy, University of Virginia, Charlottesville, VA 22904, USA\\
$^2$ Steward Observatory, University of Arizona, 933 N. Cherry Ave., Tucson, AZ 85721, USA  \\ 
$^3$ Gemini Observatory, 670 N. Aohoku Place, Hilo, HI 96720, USA\\
$^4$ Department of Astronomy, University of California, Berkeley, CA 94720-3411, USA \\
$^5$ Department of Astronomy \& Astrophysics, University of California, San Diego, 9500 Gilman Drive, MC 0424, La Jolla, CA 92093-0424, USA \\
$^6$ IPAC, MC 100-22, Caltech, 1200 E.\ California Blvd., Pasadena, CA 91125, USA \\}

\date{Accepted XXX. Received YYY; in original form ZZZ}

\pubyear{\the\year{2024}}

\begin{document}
\pagerange{\pageref{firstpage}--\pageref{lastpage}} \pubyear{2024}
\maketitle
\label{firstpage}

\begin{abstract}
We present spectra of the supernova (SN) impostor AT~2016blu spanning over a decade. This transient exhibits quasiperiodic outbursts with a $\sim$113~d period, likely triggered by periastron encounters in an eccentric binary system where the primary star is a luminous blue variable (LBV). The overall spectrum remains fairly consistent during quiescence and eruptions, with subtle changes in line-profile shapes and other details. Some narrow emission features indicate contamination from a nearby H~{\sc ii} region in the host galaxy, NGC~4559. Broader H$\alpha$ profiles exhibit Lorentzian shapes with full width at half-maximum intensity (FWHM) values that vary significantly, showing no correlation with photometric outbursts or the 113~d phase. At some epochs, H$\alpha$ exhibits asymmetric profiles with a stronger redshifted wing, while broad and sometimes multicomponent P~Cygni absorption features occasionally appear, but are again uncorrelated with brightness or phase.  These P~Cygni absorptions have high velocities compared to the FWHM of the H$\alpha$ emission line, perhaps suggesting that the absorption component is not in the LBV's wind, but is instead associated with a companion. The lack of phase dependence in line-profile changes may point to interaction between a companion and a variable or inhomogeneous primary wind, in an orbit with only mild eccentricity. Recent photometric data indicate that AT~2016blu experienced its \nth{21} outburst around 2023 May/June, as predicted based on its period. This type of quasiperiodic LBV remains poorly understood, but its spectra and erratic light curve resemble some pre-SN outbursts like those of SN~2009ip.

\end{abstract}

\begin{keywords}
stars: individual: AT~2016blu -- stars: massive -- stars: variables: general -- galaxies: NGC~4559.

\end{keywords}

\section{INTRODUCTION}\label{sec:intro}
AT~2016blu (also known as NGC~4559OT) is an extragalactic supernova (SN) impostor located in an outer spiral arm of its host galaxy NGC~4559, at a distance of $8.91 \pm 0.3$~Mpc \citep{Mc17}. AT~2016blu was discovered in 2012 during the Lick Observatory Supernova Search \citep[LOSS;][]{L02,F01,K12} with the 0.76~m Katzman Automatic Imaging Telescope (KAIT). In a previous paper, we analyzed the photometric evolution of AT~2016blu \citep[hereafter Paper I;][]{A16blu}, revealing at least 19 recorded outbursts, recurring every $113 \pm 2$~d.  Its outbursts have a typical brightness of $m_R \approx 17$ mag, corresponding to an absolute magnitude of approximately $-$12.8 with distance and reddening corrections. We proposed that these outbursts are caused by periastron encounters in an interacting eccentric binary system \citep{K10,SN11}, in which the primary star is a luminous blue variable (LBV). 

AT~2016blu is not the only SN impostor known for its quasiperiodic outbursts. SN~2000ch in NGC~3432 experienced multiple outbursts from 2000 to 2009, as documented by \cite{W04} and \cite{P10}. Continuing this pattern, SN~2000ch has undergone a total of 23 brief outbursts since 2000 \citep{A00ch}. These outbursts also repeat every $200.7 \pm 2$~d, which hints at a possible binary nature \citep{P10,SN11,S11,A00ch}, similar to AT~2016blu. In both systems, the variability of outbursts can be influenced by the intrinsic S Doradus-like variability of the primary star \citep{vg01,G09}, which may cause the binary interaction to be somewhat different at each periastron pass, with some periastron encounters leading to photometric outbursts that are weaker or even absent \citep{A00ch,A16blu}.

Both AT~2016blu and SN~2000ch have been noted for their resemblance to $\eta$ Carinae. Although they are not identical to $\eta$ Carinae, they exhibit similar behaviour, suggesting similar underlying physical processes. Their brief outbursts are reminiscent of the peaks observed in $\eta$ Carinae in the few years leading up to its 19th century Great Eruption, which occurred during periastron in its binary system \citep{sf11,SN11}. The recurring outbursts of SN~2000ch and AT~2016blu also resemble the repeating peaks of SN~2009ip before its SN explosion \citep{S10,P13}. Notably, the LBV-like outbursts of SN~2009ip culminated in a luminous Type IIn SN-like event in 2012 \citep{M13,SMP14,S22}. This event provides compelling evidence that LBVs can be progenitors of Type IIn supernovae, contrary to most theoretical predictions \citep{M13,S14araa}.  Similar LBV-like variability before an SN was also seen in SN~2015bh \citep{E16,B18}. However, not all Type IIn SNe originate from LBV-like progenitors. Some may arise from other massive stars, such as red supergiants \citep[RSGs;][]{F02,S09}, indicating diverse evolutionary pathways. Likewise, not all LBVs necessarily end as Type IIn supernovae. LBV-like outbursts have also been observed before Type Ibn events, such as SN~2006jc \citep{F07,P07}, and models suggest that they may also lead to Type IIb SNe
\citep[e.g., SN~2008ax,][]{Groh13}. Additionally, in traditional single-star evolutionary models, strong LBV mass loss drives the evolution to Wolf-Rayet (WR) stars that can ultimately explode as Type Ibc SNe.

Given the similarities in photometric evolution between AT~2016blu, SN~2000ch, and other well-known LBVs such as $\eta$ Carinae and SN~2009ip, it is pertinent to study their spectral evolution as well. LBVs, being highly variable stars, exhibit significant spectral changes over time. Traditional expectations are that during outbursts, they resemble cool supergiants with spectral types A or F, and during the quiescent state, they appear as hotter B-type supergiants or Ofpe/WN9 stars \citep[see][for reviews]{H94,smith17,M24}. LBVs in cool outburst states typically exhibit a temperature around 8000~K, and display a combination of absorption, emission, and P Cygni lines, with line widths suggesting relatively slow outflow speeds between $100$ and $500$~km~s$^{-1}$. This is not always true, however, as a subset of LBVs has been shown to brighten at relatively constant (hot) temperatures \citep{S11,smith20}. This class of hot LBV outbursts includes the precursor of SN~2009ip \citep[pre-SN outbursts in 2009–2011;][]{S10}, SN~2000ch \citep{W04,S11}, HD~5980 \citep{K04}, and both MCA-1B \citep{smith20} and Romano's star \citep{P16} in M33.   SN~2009ip, for instance, exhibited strong, narrow emission lines with a speed of about $550$~km~s$^{-1}$ during its pre-SN outbursts, similar to that observed in other SN impostors \citep{V07,S11,V12a}, but it maintained a blue continuum slope and He~{\sc i} emission lines at all phases \citep{S10}. Some lines in SN~2009ip also showed weak, broad, and blueshifted absorption wings reaching high velocities up to $\sim 10,000$~km~s$^{-1}$ \citep{S10,P13}. Similarly, $\eta$ Carinae primarily shows eruptive outflows at speeds around $600$~km~s$^{-1}$ or less, but about 1\% of its material reaches much higher speeds of 5000 to 10,000~km~s$^{-1}$ \citep{S08,S18,S18b}.

\cite{W04}, \cite{P10}, and \citet{S11} also studied the optical spectra of SN~2000ch, revealing evolving spectral characteristics. \cite{W04} observed changes in Balmer emission lines and the absence of typical LBV spectral features like Fe {\sc ii} lines, consistent with maintaining a hotter temperature in outburst, as noted above. \cite{P10} reported significant changes in spectral features, including increased H$\alpha$ width, stronger He {\sc i} lines, and the emergence of P Cygni profiles. These observations also showed some exceptionally high terminal wind velocities, unusual for known LBV winds. Overall, SN~2000ch's spectral features varied, but resembled those of some SN impostors like the precursor of SN~2009ip and H-rich WR stars.

Studying objects like SN~2000ch and AT~2016blu will not only uncover the physical mechanisms driving quasiperiodic eruptions in massive stars, but may also provide critical insights into their post-main-sequence evolution, including the diverse population of LBVs and the progenitors of interacting SNe. Moreover, investigating these systems helps reconcile discrepancies between theoretical models and observations, refining our understanding of massive star evolution.

In this paper, we analyze the spectroscopic evolution of AT~2016blu and explore its similarities with other SN impostors. Section~\ref{sec:obs} presents the latest photometric and spectroscopic observations. In Section~\ref{sec:LC}, we document the new outbursts observed since the publication of Paper I until 2024. Section~\ref{sec:spectra} discusses the spectral evolution of various lines, and Section~\ref{sec:fwhm} covers the H$\alpha$ emission-line profiles, including the estimation of full width at half-maximum intensity (FWHM) values. Section~\ref{sec:comparison} focuses on comparing the spectra of AT~2016blu with those of SN~2000ch and the precursor of SN~2009ip. We discuss our findings in Section~\ref{sec:Discussion}, and conclude with a summary in Section~\ref{sec:conclusion}.

\section{OBSERVATIONS}\label{sec:obs}
\subsection{New Photometry}\label{sec:obsphot}
\begin{figure*}
  \includegraphics[width=\textwidth]{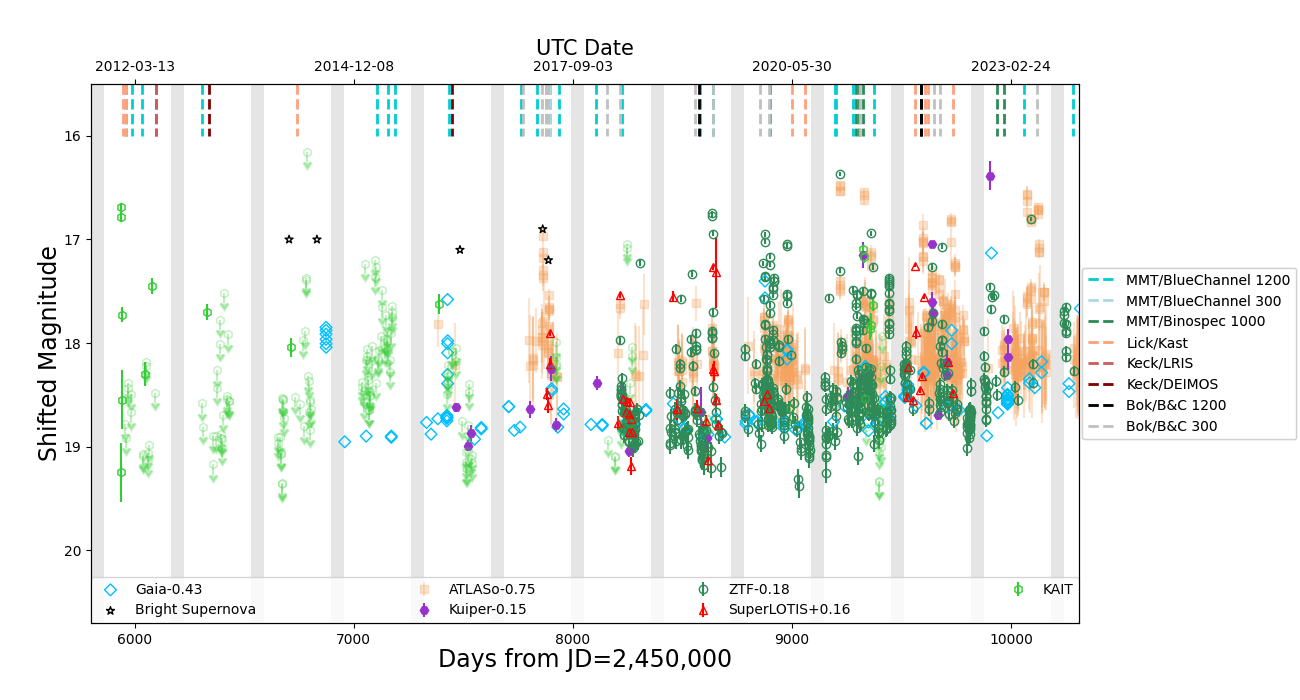}
  \label{fig:AllLCwshift}
\caption{The updated photometric evolution of AT~2016blu until 2024 January. Epochs of spectroscopic observations are marked with vertical ticks. Grey vertical sections highlight periods each year, from August 26 to October 24, when observing AT~2016blu is challenging owing to its proximity to the Sun in the sky. For a closer look at recent outbursts, see Fig.~\ref{fig:RecentZoomed}.}\label{fig:AllLC}
\end{figure*}

\begin{figure}
\subfloat[]{
  \includegraphics[width=0.9\columnwidth]{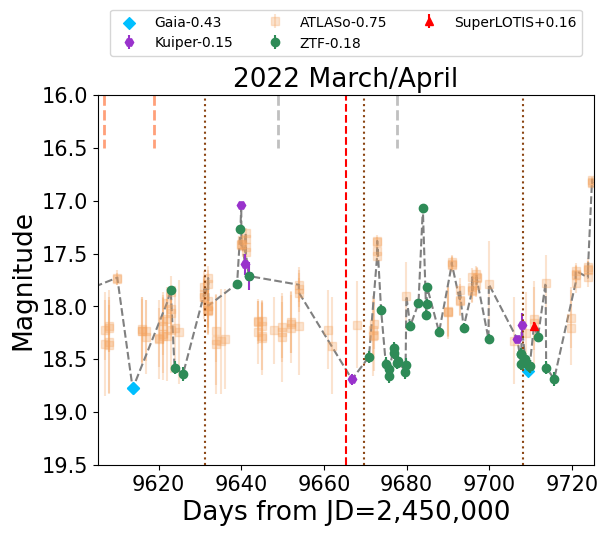}
  \label{fig:RecentZoomed1}
}
\newline
\subfloat[]{
  \includegraphics[width=0.9\columnwidth]{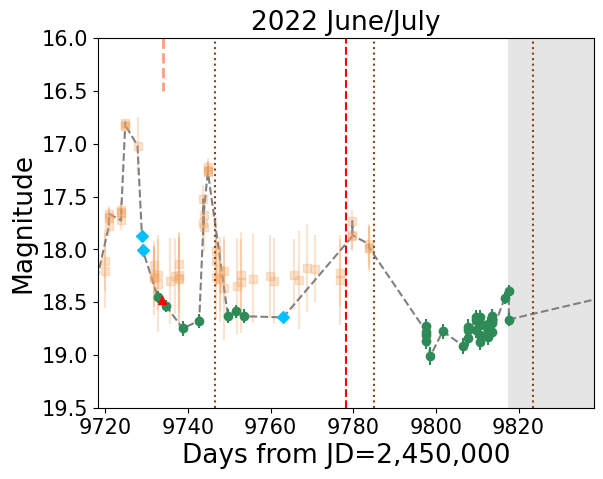}
  \label{fig:RecentZoomed1a}
}
\newline
\subfloat[]{
  \includegraphics[width=0.9\columnwidth]{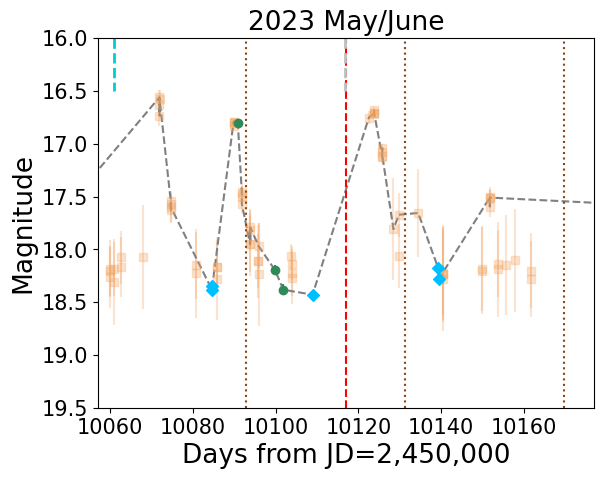}
  \label{fig:RecentZoomed2}
}
\caption{Zoomed-in views of the light curve of AT~2016blu during its recent outbursts, with the timing of each event indicated at the top of the frame. The red dashed line marks the reference epoch, based on the most probable period of about 113~d, linked to the first 2012 January outburst. The brown dotted line represents the reference epoch using the second significant peak identified in the period analysis, around 38.5~d \citep{A16blu}.}\label{fig:RecentZoomed}
\end{figure}

Paper I presents the optical and infrared photometry of AT~2016blu from 1999 to 2022. This work includes updated optical photometry of AT~2016blu using the latest data from multiple sources: the {\it Gaia} space telescope \citep{G16}, the Zwicky Transient Facility \citep[ZTF;][]{B19} public survey, and the ATLAS Project \citep[Asteroid Terrestrial-impact Last Alert System;][]{T18}. For detailed information about photometry, refer to Paper I. Fig.~\ref{fig:AllLC} shows the updated light curve, and Fig.~\ref{fig:RecentZoomed} displays the zoomed-in light curves for the recent outbursts (see Section~\ref{sec:LC} for more details).

\subsection{Spectroscopy}\label{sec:obsspec}
\begin{figure*}
\subfloat[]{
  \includegraphics[width=0.7\textwidth]{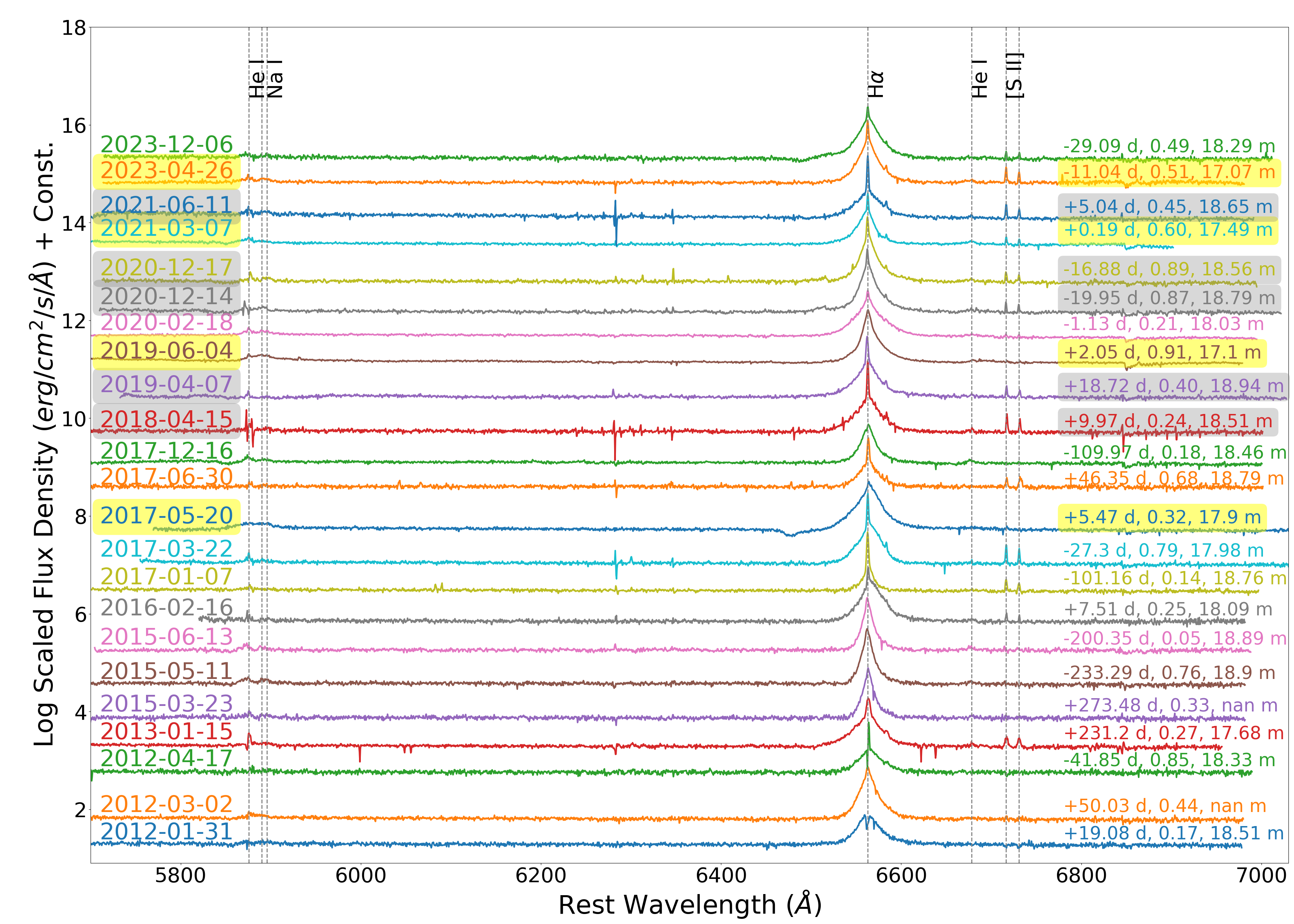}
  \label{fig:BCH1200}}\\
\subfloat[]{
  \includegraphics[width=0.7\textwidth]{
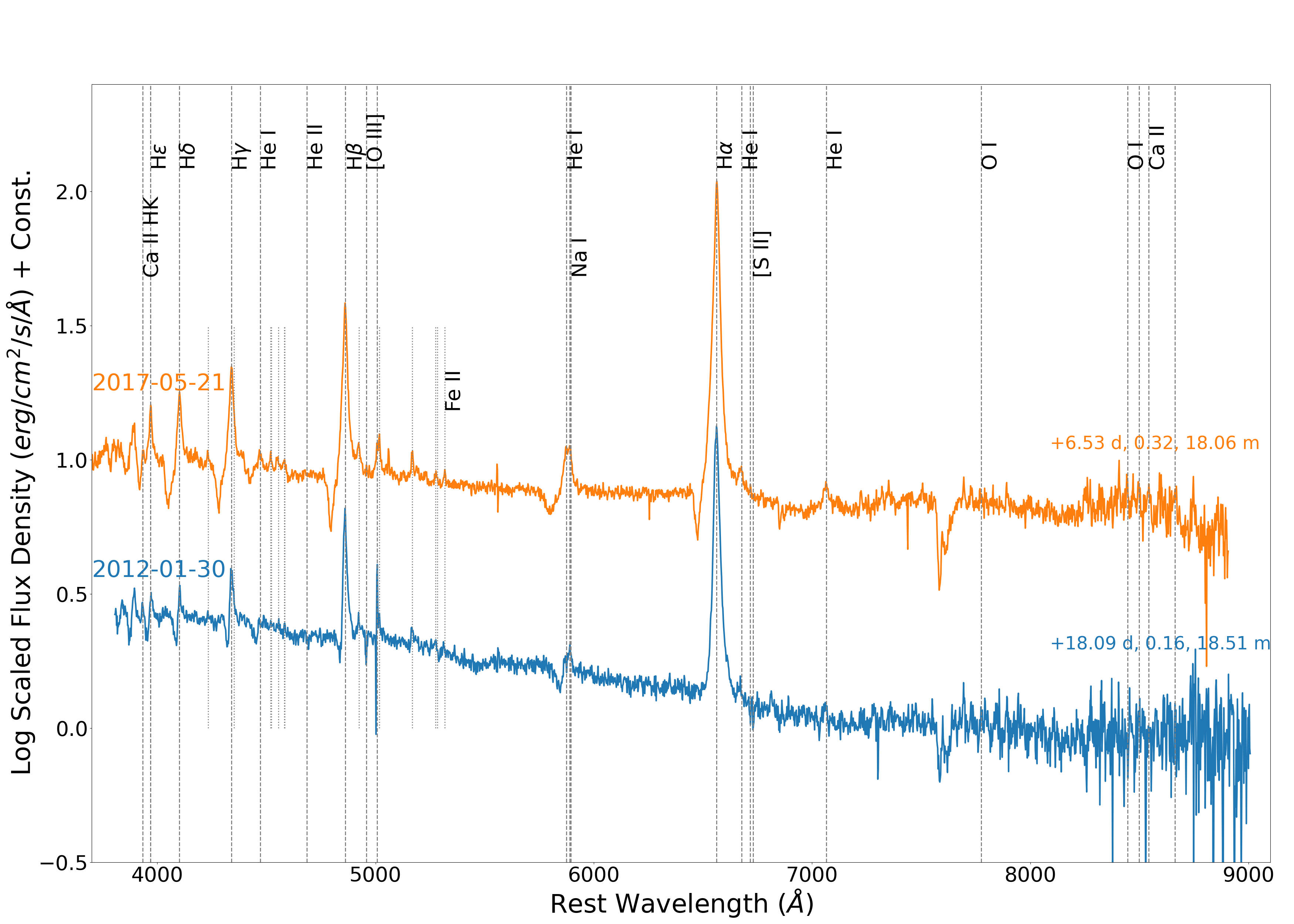}
  \label{fig:BCH300}}
\caption{Optical spectra of AT 2016blu obtained with MMT/BlueChannel using gratings with 1200 lines mm$^{-1}$ (top panel) and 300 lines mm$^{-1}$ (bottom panel). Note that all spectra throughout the paper have been corrected for the adopted host-galaxy redshift of $z=0.00261$. Spectra corresponding to times when AT~2016blu had an interpolated magnitude brighter than 18 and are taken within 15 days of peak brightness are annotated in yellow. Spectra most likely taken during the quiescent state are annotated in grey. The three values on the right side show the days after or before the documented peak brightness, the 113~d period phase, and the interpolated magnitude at the time the spectrum was taken. Balmer, He~{\sc i}, and Fe~{\sc ii} lines, as well as P~Cygni profiles, are visible.}\label{fig:BCH}
\end{figure*}

\begin{figure*}
\subfloat[]{
  \includegraphics[width=0.78\textwidth]{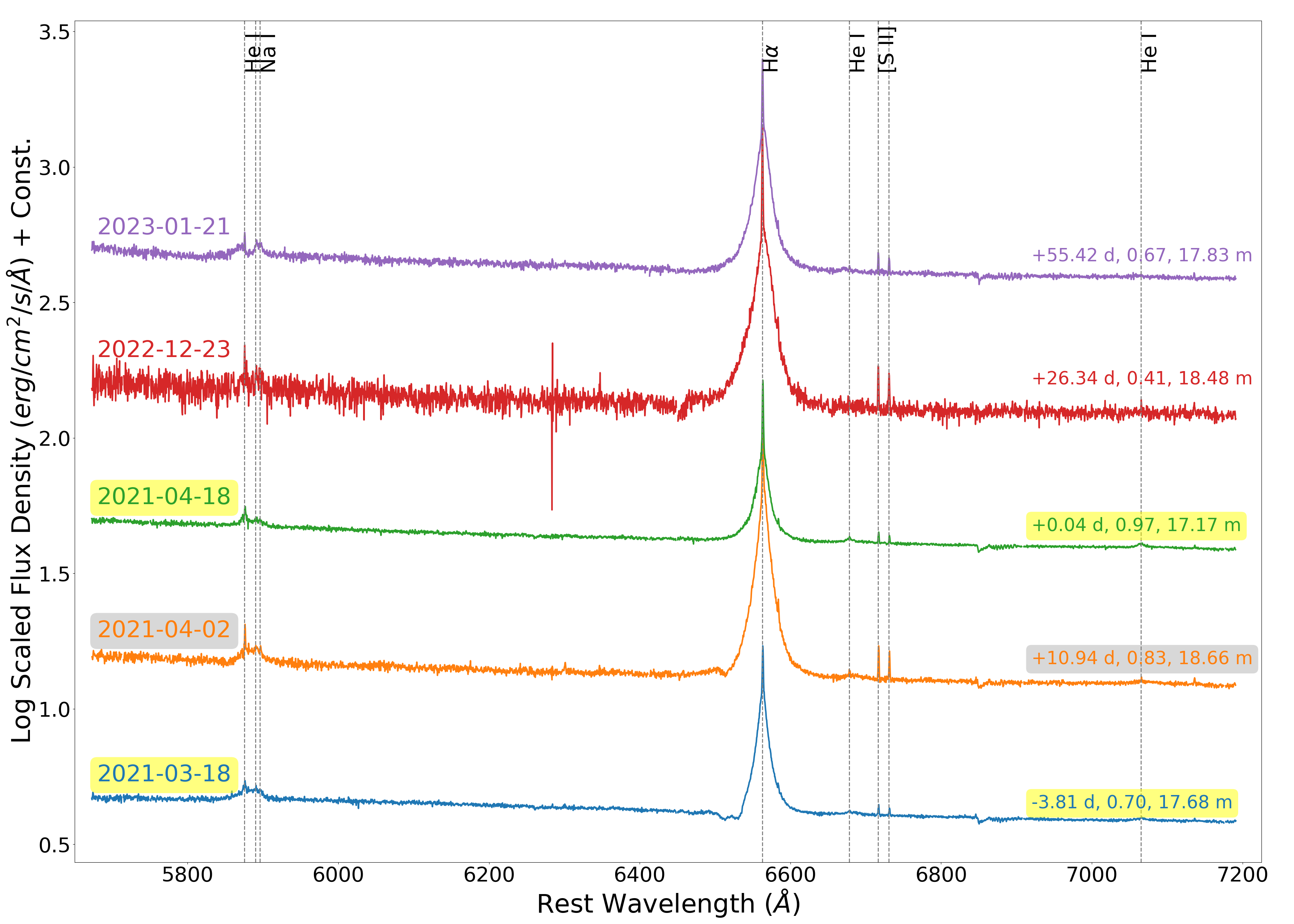}
  \label{fig:Bino}}\\
\subfloat[]{
  \includegraphics[width=0.78\textwidth]{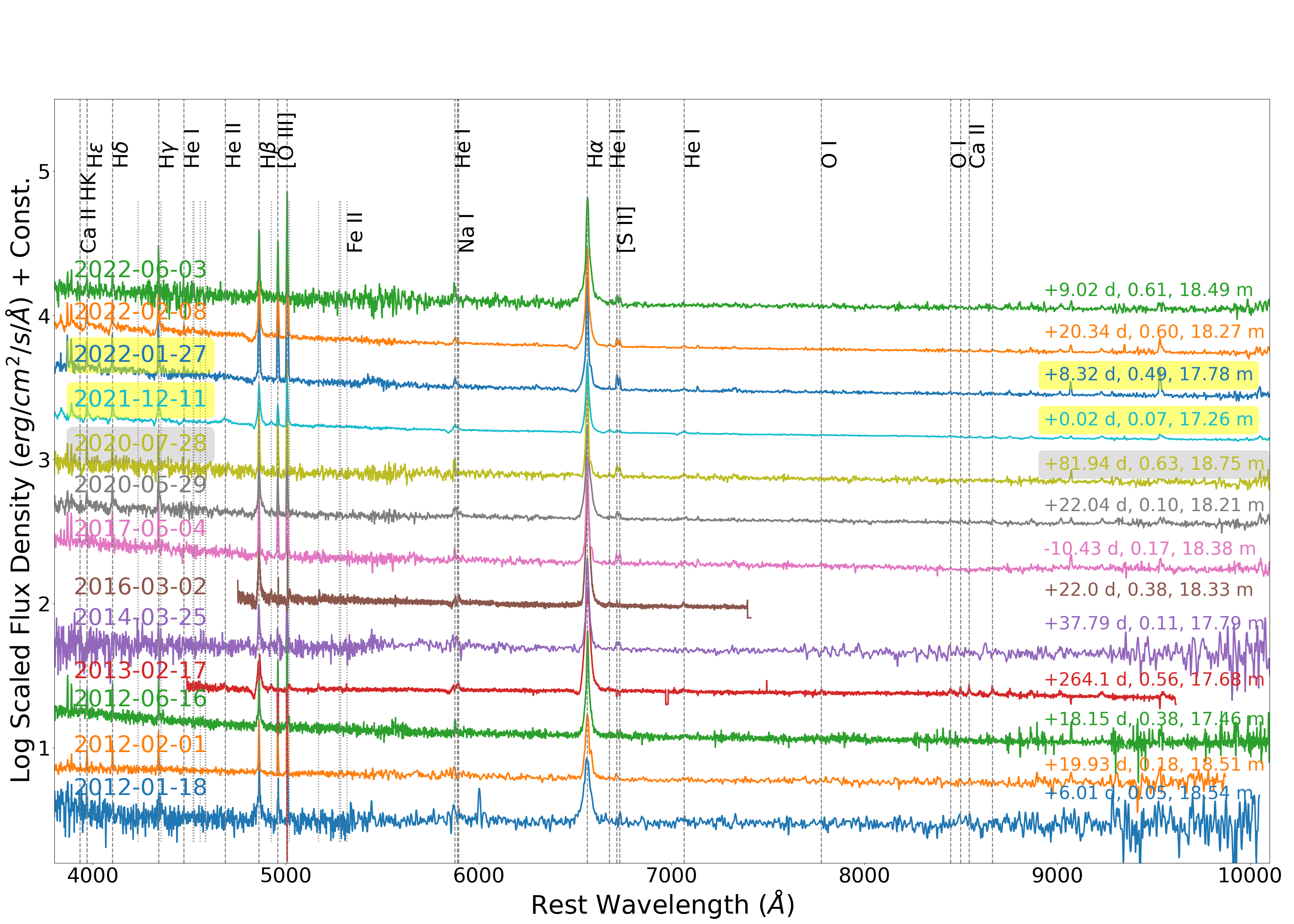}
  \label{fig:UCB}}
\caption{The top panel shows the spectra of AT~2016blu from MMT/Binospec. The Binospec data reveal narrow H$\alpha$ emission from the underlying H~{\sc ii} region or CSM. Binospec spectra also display He~{\sc i} lines during both outburst and quiescent states. The bottom panel shows a series of high-, moderate- and low-resolution spectra of AT~2016blu from Keck/LRIS, Keck/DEIMOS, and Lick/Kast. Some spectra show O~{\sc i} and Ca~{\sc ii} lines.}\label{fig:ucbbino}
\end{figure*}

%mmt blue 1200 and 300 and bino
We obtained optical spectra of AT~2016blu using the BlueChannel and Binospec \citep{F19} Spectrographs on the 6.5~m Multiple Mirror Telescope (MMT). Observations with the MMT BlueChannel were performed using 1200 and 300 lines mm$^{-1}$ gratings, with approximate resolutions of 1.5 and 6.5~\AA\ (roughly 70 and 300 km s$^{-1}$, respectively). 
%\MA{\href{xc}{\MA{Page 3 of this document}}}. 
Meanwhile, the MMT Binospec observations utilised a 1000 lines mm$^{-1}$ grating with a nominal spectral resolution $R = \lambda$/$\Delta\lambda$ of about 3900 ($\sim$ 80 km s$^{-1}$), although some of our spectra had somewhat better resolution on nights with good seeing. 
%\MA{ \href{http://mingus.mmto.arizona.edu/~bjw/mmt/binospec_info.html}{\MA{Table 1 of this document}}. But when i estiamte the FWHM of [S II] lines, they vary between 50 to 72 ish km/s. }. 
Reductions for the MMT BlueChannel were carried out using IRAF\footnote{IRAF is distributed by the National Optical Astronomy Observatory, which is operated by the Association of Universities for Research in Astronomy (AURA) under a cooperative agreement with the U.S. National Science Foundation.}. Flux calibration was achieved using spectrophotometric standards at similar airmasses, taken during the same night as the observation. The MMT Binospec data were reduced using the Binospec pipeline \citep{K19}. All MMT spectra are plotted in Figs.~\ref{fig:BCH1200}, \ref{fig:BCH300}, and \ref{fig:Bino}. 

 \begin{figure*}
\subfloat[]{
  \includegraphics[width=0.78\textwidth]{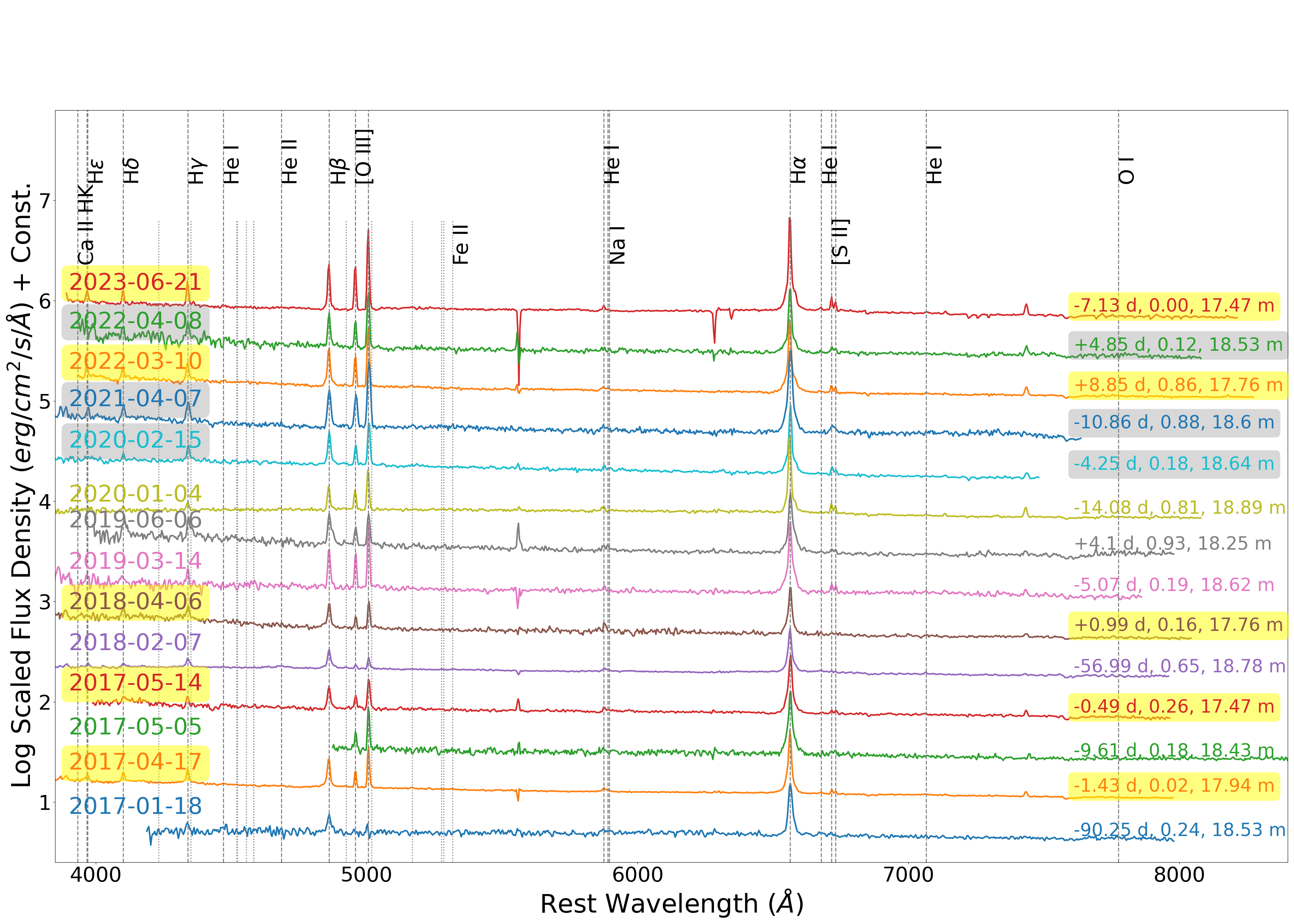}
  \label{fig:bok300}}\\
\subfloat[]{
  \includegraphics[width=0.78\textwidth]{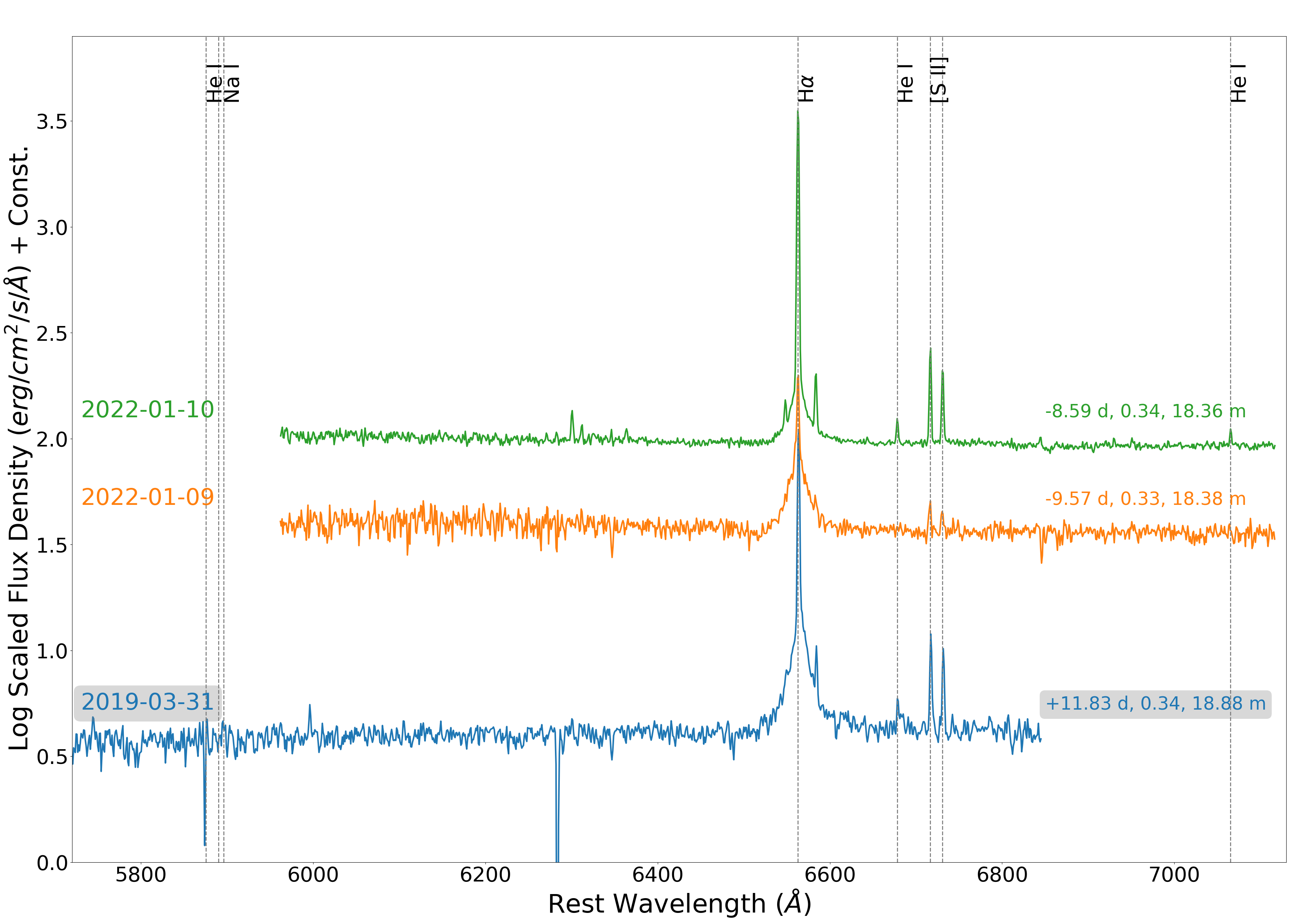}
  \label{fig:bok1200}}
\caption{Optical spectra of AT~2016blu from Bok B\&C. The top and bottom panels show the 300 and 1200 lines mm$^{-1}$ gratings, respectively.  The features near 5560 and 6282~\AA\  are artifacts of the data-reduction process. Similar to the Binospec spectra, contamination from a nearby H~{\sc ii} region is evident in the bottom panel.
}\label{fig:bokall}
\end{figure*}

%Keck and Lick
Two spectra were obtained at the W. M. Keck Observatory using the Deep Imaging Multi-Object Spectrograph \citep[DEIMOS;][]{F03} with  $R \approx 4400$ ($\sim$ 70 km s$^{-1}$),
%\MA{Table 7 of https://academic.oup.com/mnras/article/415/1/773/989885}. 
and one spectrum was collected at Keck using the Low-Resolution Imaging Spectrometer \citep[LRIS;][]{O95}, with $R \approx 1000$ ($\sim 300$ km s$^{-1}$).
%\MA{Table 7 of https://academic.oup.com/mnras/article/415/1/773/989885} at the W. M. Keck Observatory. 
Additionally, spectra were acquired using the Kast double spectrograph on the Lick 3~m Shane reflector with $R \approx 700$ ($\sim 430$ km s$^{-1}$).
%\MA{Table 7 of https://academic.oup.com/mnras/article/415/1/773/989885}. 
These Keck and Lick spectra were reduced using standard techniques, including bias subtraction, flat-fielding, local sky subtraction, and extraction of one-dimensional spectra. Fig.~\ref{fig:UCB} presents all the spectra from the Keck and Lick Observatories.

The DEIMOS spectrum from 2013-02-17 and MMT BlueChannel spectra from 2012-01-31 and 2012-04-17 exhibit narrow absorption features superimposed on emission lines. These are likely due to oversubtraction of the background sky and surrounding H~{\sc ii} region emission during data reduction; they are thus artifacts,
%These features are most likely artifacts, 
as similar features are not observed at any other time.

%Bok
Optical spectra were also obtained using the Boller \& Chivens (B\&C) spectrograph on the 2.3 m Bok telescope, utilising both 1200 and 300 lines mm$^{-1}$ gratings with approximate resolutions of 1.8 
%\MA{6562.79/3600, 3600 comes from https://academic.oup.com/mnras/article/430/3/1801/979371} 
and 8 
%\MA{\href{https://repository.arizona.edu/bitstream/handle/10150/666597/azu_etd_hr_2021_0074_sip1_m.pdf?sequence=1&isAllowed=y}{\MA{Page 3 of this document}}} 
\AA \ ($\sim 80$ and 370 km s$^{-1}$), respectively. Data reduction for these spectra was performed following standard IRAF procedures. The Bok spectra are displayed in Figs.~\ref{fig:bok300}, and \ref{fig:bok1200}. 

All reduced spectra have been corrected for the redshift of AT~2016blu's host, which is $z = 0.00261$. The UTC dates of observation are listed next to each spectrum.

\section{Updated Light curve}\label{sec:LC}
Fig.~\ref{fig:AllLC} shows the most updated light curve for AT~2016blu from 2012 until 2024. The solid vertical grey bands mark the time interval each year, from August 26 to October 24, during which AT~2016blu is challenging or impossible to observe due to its proximity to the Sun in the sky.  All the spectroscopic data are marked with dashed vertical lines in the light curve.  Photometric data from ZTF are in the $r$ band, {\it Gaia} in the $G$ band, and ATLAS in the $o$ band. 
Therefore, we have adjusted the fluxes from each dataset to approximately match the base quiescent magnitudes across all datasets. Recent photometric observations reveal that AT~2016blu experienced another peak in 2022 April. This peak was not covered in the data available at the time of the first paper's publication. We also identify a series of additional ATLAS peaks around mid-2022 that were previously overlooked. Furthermore, we have observed that AT~2016blu experienced an outburst in 2023 as predicted in Paper I. Fig.~\ref{fig:RecentZoomed} shows a zoomed-in version of the light curve for each event.

In this section, we provide a detailed description of each event. The designation for each event corresponds to the date of the two brightest peaks. According to Paper I, the outbursts appear to recur with a likely periodicity of $\sim 113$~d. As explained in Paper I, the bright eruption events are very brief (a few days or weeks), and instead of one eruption landing at the expected time for the periodicity (given by the red vertical line in Fig.~\ref{fig:RecentZoomed}), we see groups of several brightening events clustered around the predicted period.  According to the scenario outlined in Paper I, this may arise from a companion interacting with a clumpy wind or series of shells ejected irregularly by the LBV, with a tendency to occur more often near periastron, but not exactly at the time  of periastron. Additionally, the period analysis revealed a lower amplitude peak roughly at 39~d, which is probably related to multiple spikes occurring either before or after each outburst (see Paper I). Each plot features a red dashed line indicating the expected time of the outburst, which is derived from $d_{\rm red} = 2,455,938.9 + 113\,n$. A brown dotted line in each plot marks the reference epoch, determined using the second most dominant period found in the periodogram.

{\it 2022 March/April:} In Paper I, we documented the 18th outburst of AT~2016blu. The peak observed was more consistent with the shorter period, as indicated by a brown dotted line, rather than the longer period marked by a red dashed line. New data from ZTF DR23 suggest that, similar to many of AT~2016blu’s previously documented outbursts, this event also exhibited multiple peaks of approximately similar magnitudes; see Fig.~\ref{fig:RecentZoomed1} for details.

{\it 2022 June/July:} Fig.~\ref{fig:RecentZoomed1a} shows that AT~2016blu experienced a series of peaks in mid-2022. ATLAS data indicate that in 2022 May, AT~2016blu reached a magnitude of 16.8, and shortly after, in June, it brightened again to 17.2. The second peak is consistent with the shorter $\sim$39 d period marked by the brown dotted line. Around the expected time of the outburst for the main 113 d period, in July, it brightened once more, though not as bright as the previous peaks.

{\it 2023 May/June:} Based on the detected period of approximately 113~d, it was predicted that AT~2016blu would experience an outburst in 2023 February/March, and then again in late June. Unfortunately, there is still insufficient public data to confirm whether an outburst occurred in 2023 February/March. However, the predicted June 2023 outburst was observed. Fig.~\ref{fig:RecentZoomed2} shows that AT~2016blu underwent its \nth{21} recorded outburst, reaching an apparent magnitude of 16.6 starting in early May of 2023, with multiple subsequent peaks continuing to late June. ATLAS data show a peak in June that aligns with the predicted time of outburst. \cite{D23} also monitored AT~2016blu around the proposed time and reported an outburst of AT~2016blu in late June, reinforcing the ATLAS data.

\section{Overview of the spectra}\label{sec:spectra}
The spectra of AT~2016blu are shown in Figs.~\ref{fig:BCH}--\ref{fig:bokall}. Each spectrum is labeled with the UTC date of observation, the number of days relative to a documented peak in the light curve, the orbital phase using a period of 113~d, and interpolated magnitudes at the time the spectrum was observed. We estimated the system's approximate $R$ magnitude at the time of the spectroscopic observations using all available photometry, excluding faint ATLAS data ($>$18 mag), which are noisy due to AT~2016blu's brightness being comparable to the telescope's limiting magnitude. 
Spectra with ``nan'' magnitudes indicate that no magnitude could be interpolated owing to the absence of photometric data within 30 days of the spectroscopic observation. Given the brief and erratic nature of AT~2016blu's outbursts, any future data releases, which may include older, previously unreleased photometry, could lead to updated magnitudes. Spectra that have an interpolated magnitude brighter than 18 and are captured within 15 days of peak brightness in the light curve are highlighted in yellow, while those taken during quiescent states are shaded grey. For spectra not highlighted, insufficient photometry makes it difficult to determine the state of the system at the time of the spectroscopic observation. All spectra are scaled using the continuum portion of each spectrum.

AT~2016blu's spectra are dominated by strong hydrogen Balmer emission lines with narrow peaks and broadened wings. Various lines are identified across all spectra. However, not all lines are visible in every epoch. Notably, a handful of spectra (e.g., MMT/BlueChannel 300 lines mm$^{-1}$ grating) show weak Fe~{\sc ii} lines, but the majority of the spectra lack the Fe~{\sc ii} and [Fe~{\sc ii}] emission lines that are commonly observed in the spectra of LBVs \citep{H06,M17}. Some spectra also show Ca~{\sc ii}, and O~{\sc i} lines. However, the absence of strong Fe~{\sc ii} and Ca~{\sc ii} in most spectra and the presence of He~{\sc i} emission lines suggest that AT~2016blu is significantly hotter than a typical LBV in its eruptive state, as noted above.  
While P~Cygni profiles are occasionally evident, some spectra do not display the characteristic P~Cygni profiles typical of LBVs in outburst. Despite some subtle changes in details, such as the width of H$\alpha$ and the variable P~Cygni absorption, the overall optical spectrum of AT~2016blu remains fairly consistent throughout both the outburst and quiescent states.

[S~{\sc ii}] and strong [O~{\sc iii}] lines have also been detected, suggesting potential contamination from a neighbouring H~{\sc ii} region. This is supported by a narrow component observed overlaying the broader H$\alpha$ emission base, as shown in Figs.~\ref{fig:Bino} and  \ref{fig:bok1200}, although some of the narrow H$\alpha$ could also originate from circumstellar material (CSM).  It remains challenging to determine the extent of contribution from the H~{\sc ii} region where AT~2016blu is located. The Binospec pipeline selects a local sky position, while the BlueChannel data are reduced by subtracting the H~{\sc ii} region emission, leading to variations in the strength of the narrow lines originating from the H~{\sc ii} region.  Notably, during several MMT/BlueChannel epochs where the H~{\sc ii} region emissions have been effectively subtracted, we do not detect significant narrow H$\alpha$ emission in the transient. This strongly suggests that the observed narrow components primarily originate from the H~{\sc ii} region, and thus have been excluded from our FWHM analysis (see Section~\ref{sec:fwhm}). %\N{(ok, good point, arguing that the narrow Halpha is probably not CSM.)}

\subsection{Hydrogen Balmer Lines}\label{sec:Balmer}
\begin{figure}
\begin{overpic}[width=0.41\textwidth]{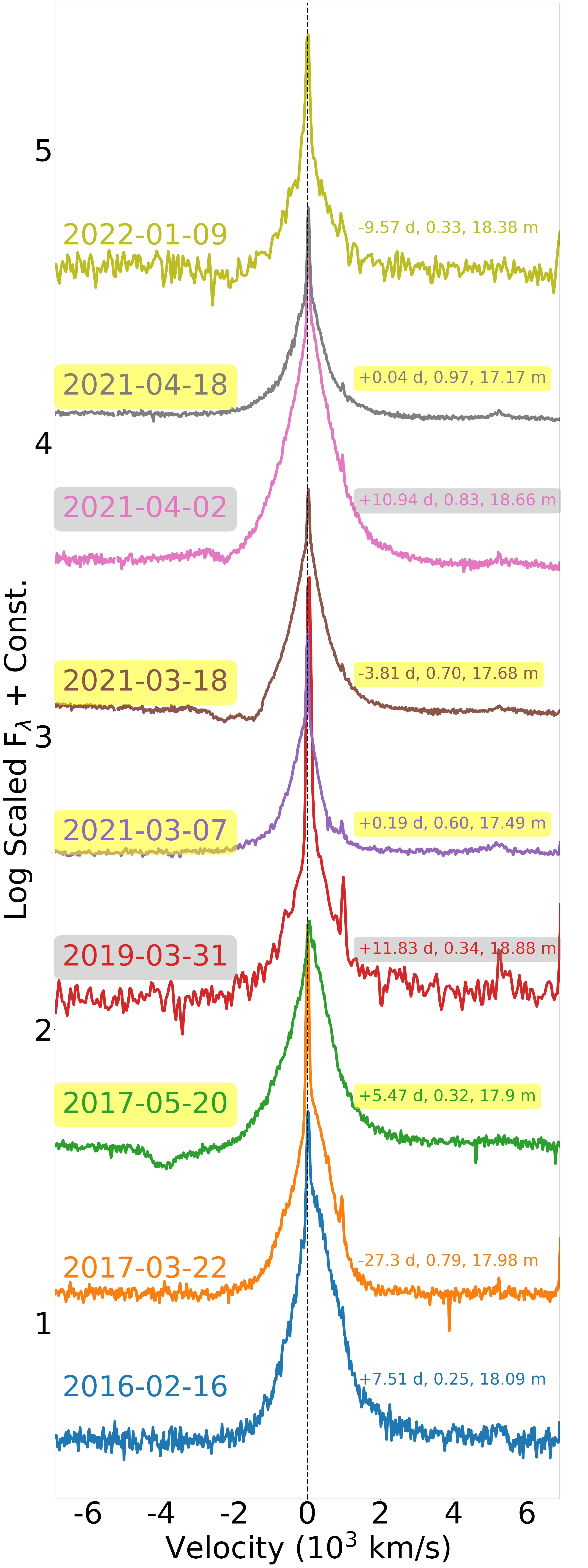}
            \put(6,97){\makebox(1,1){\textcolor{black}{\Huge H$\alpha$}}}
        \end{overpic}
\caption{Velocity profile of a subset of H$\alpha$ from high-resolution spectra. P~Cygni features are detected in both outburst and quiescent states. Some, like the 2021-03-18 spectrum, show a double-dip absorption feature. Others, such as the 2017-03-22 spectrum, also exhibit a red hump. A few, for instance the 2021-03-07 spectrum, display a blue excess.}\label{fig:halphahighres}
\end{figure}

\begin{figure}
\begin{overpic}[width=0.4\textwidth]{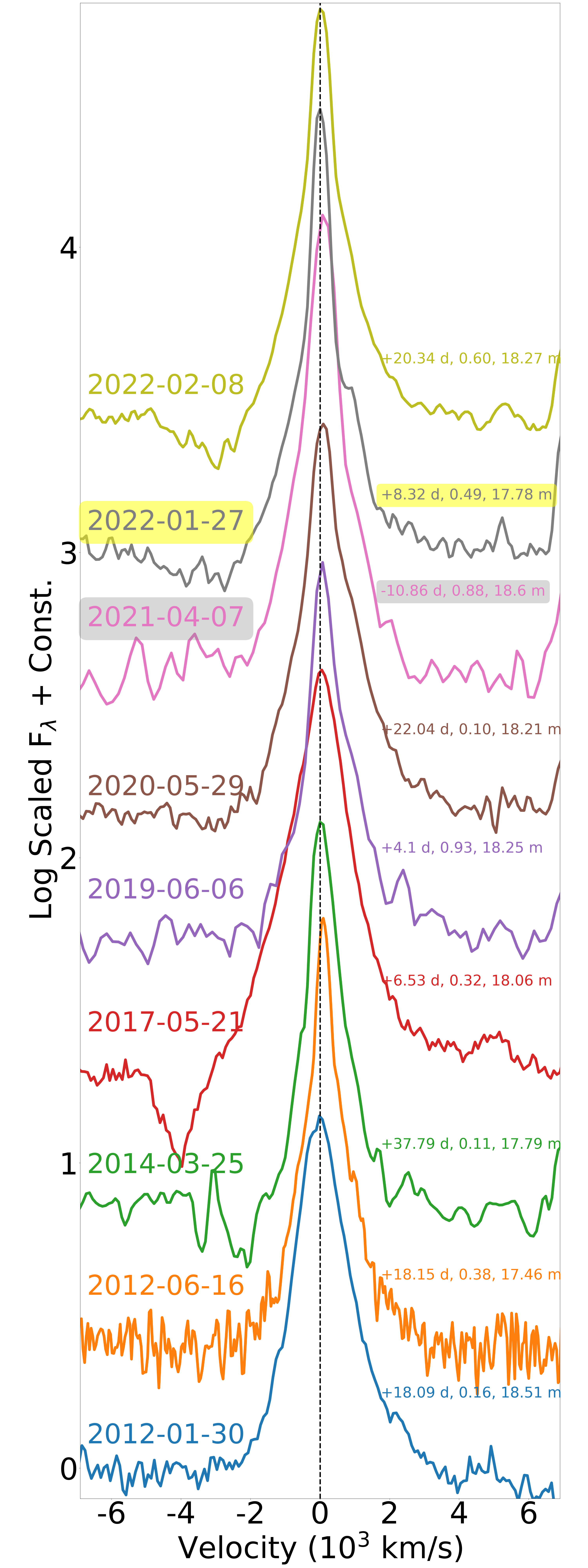}
            \put(8,97){\makebox(1,1){\textcolor{black}{\Huge H$\alpha$}}}
        \end{overpic}
\caption{Same as Fig.~\ref{fig:halphahighres}, but low-resolution data. Double-dip absorption features are visible in some spectra such as those from 2014-03-25 and 2022-01-27.}\label{fig:halphalowres}
\end{figure}

Figs.~\ref{fig:halphahighres} and \ref{fig:halphalowres} show a subset of spectra around the H$\alpha$ profile, focusing on the velocity range of $\pm$6500 km s$^{-1}$. As previously mentioned, the H$\alpha$  profile features a narrow component superimposed on a broad component, likely due to contamination from nearby H~{\sc ii} regions or CSM. These narrow components are clearly visible in high-resolution spectra, but remain unresolved in other low and mid-resolution spectra. 

In some spectra, such as those from 2016-02-16 and 2017-03-22 in the BlueChannel 1200 grating, the H$\alpha$ profile is broader and asymmetric, exhibiting a hump on the red side, or a deficit on the blue wing. In other spectra, the broad component of H$\alpha$ has a redshifted centroid, while the narrow component of the emission profile remains centred close to the H$\alpha$ rest wavelength. In Section~\ref{sec:fwhm}, we discuss radial-velocity variations across the orbital period.

Figs.~\ref{fig:halphahighres} and \ref{fig:halphalowres} also show that H$\alpha$ exhibits P~Cygni profiles at some epochs, which are typical during the outburst state of LBVs \citep{M17}. However, P~Cygni features are also observed during the quiescent state, and in some instances of outburst states of AT~2016blu no P~Cygni features are detected. The H$\alpha$ profile also displays a Lorentzian shape, and the FWHM of the H$\alpha$ profile varies over time. Detailed analyses and discussions regarding the FWHM measurements of the H$\alpha$ lines are presented in Section~\ref{sec:fwhm}.

\begin{figure}
\includegraphics[width=0.4\textwidth]{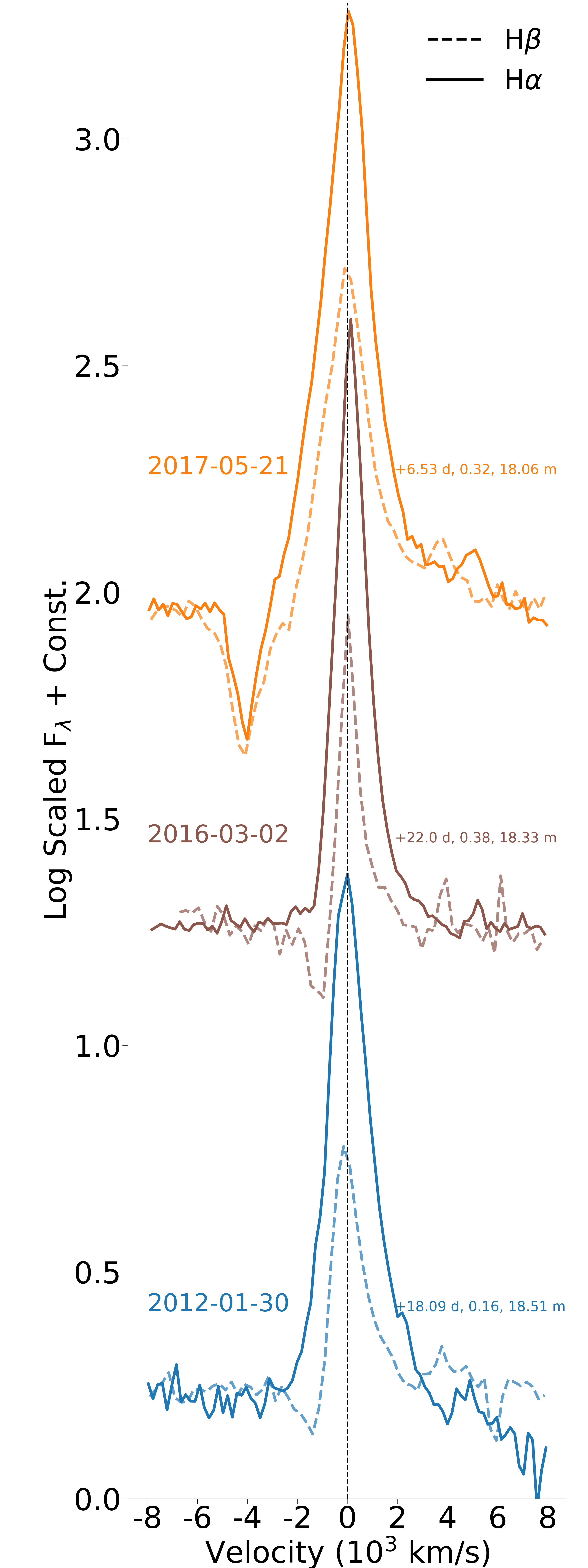}
\caption{Velocity profiles of the H$\alpha$ and H$\beta$ lines. The spectra are binned with a bin size of 4~\AA. Absorption features are generally stronger in the higher-order Balmer lines, and the P~Cygni absorption occurs at a wide range of different velocities.}\label{fig:halphabeta}
\end{figure}

Figs.~\ref{fig:BCH300}, \ref{fig:UCB}, and \ref{fig:bok300} also show higher-order Balmer emission lines. Since using a higher spectral resolution grating reduces the spectral range, the higher-order Balmer lines are only visible in low-resolution spectra. P~Cygni absorption components are also distinctly visible in some of the higher-order Balmer lines. Absorption features are generally stronger compared to those in H$\alpha$  (see Fig.~\ref{fig:halphabeta}). Similar to the spectra obtained on May 13, 2021, and March 15, 2021, by \cite{V21} and \cite{VV21}, some of our observations reveal double-dip absorption features in higher-order Balmer lines, characterised by two distinct minima within the absorption profile (e.g., 2022-01-27 and 2022-02-08, Lick spectra). The double-dip absorption feature is also visible in the H$\alpha$ line in the high-resolution Binospec spectra taken on March 18, 2021 (see Fig.\ref{fig:halphahighres}), as well as in some of the Lick spectra (see Fig.~\ref{fig:halphalowres}). The Binospec spectrum was obtained 4 days before a luminosity peak in the light curve. Interestingly, a subsequent Binospec spectrum, taken 15 days later, reveals that the double dips have merged into a single absorption feature, and later, on April 18, 2021, the P~Cygni absorption disappears completely.
There is a possibility that some double-dip features were missed owing to the low resolution of some of the spectra. \cite{V22} also reported no detection of any distinct double-dip absorption features in H$\beta$ and H$\gamma$. Double-dip absorption features are also detected in SN~2005gj \citep{T08}, but these features are not common in SN spectra. P~Cygni absorption features with multiple or discrete components are also sometimes seen in the variable winds of LBVs \citep{S83,L94,S03,S01,R11}, and these have been linked to wind variability resulting from the bistability jump \citep{G11}.  However, we note that the velocities seen here in AT~2016blu are significantly faster than the typical speeds of 100--200 km s$^{-1}$ when these are seen in normal LBVs.

\begin{figure*}
\begin{tabular}{cc}
\subfloat[Binospec 1000] {\label{fig:BinoHeI}
\includegraphics[width=0.42\linewidth]{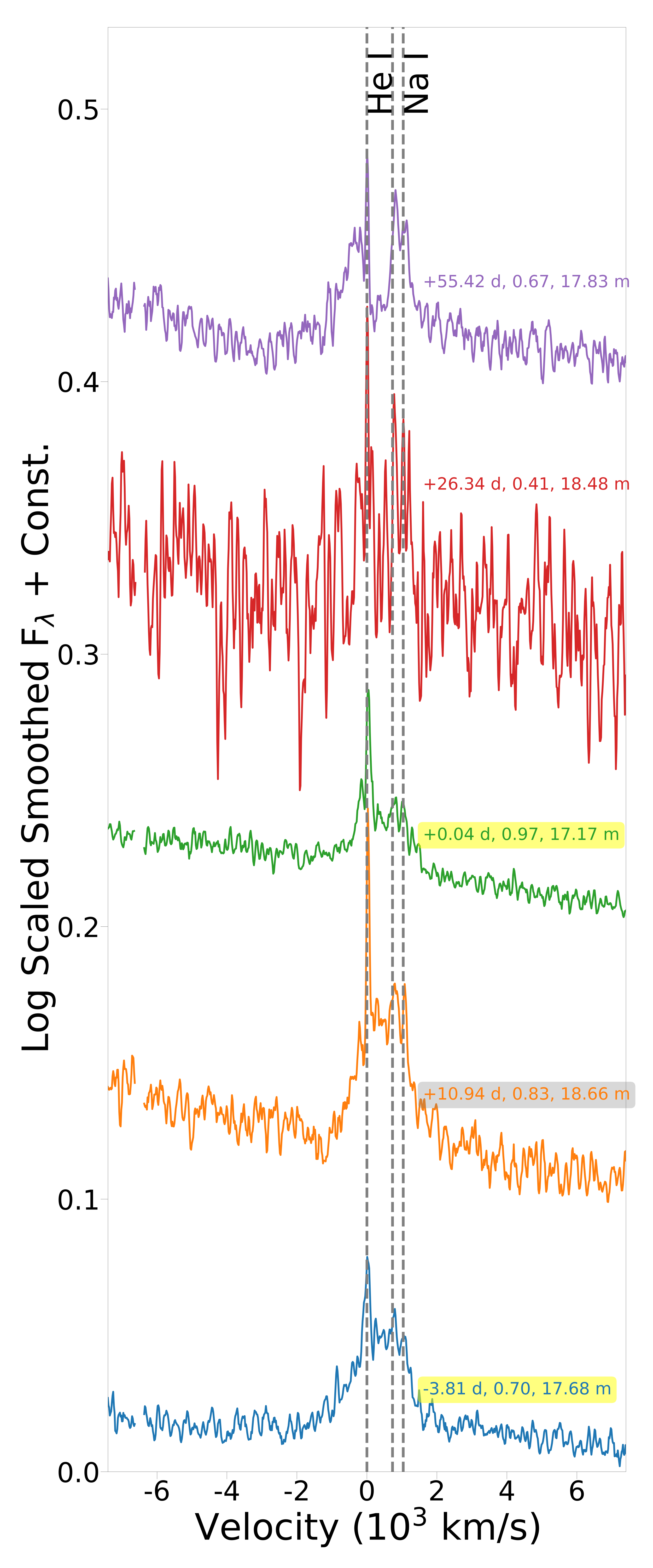}}&
    \subfloat[BlueChannel 1200]{\label{fig:BCH1200HeI}
\includegraphics[width=0.42\linewidth]{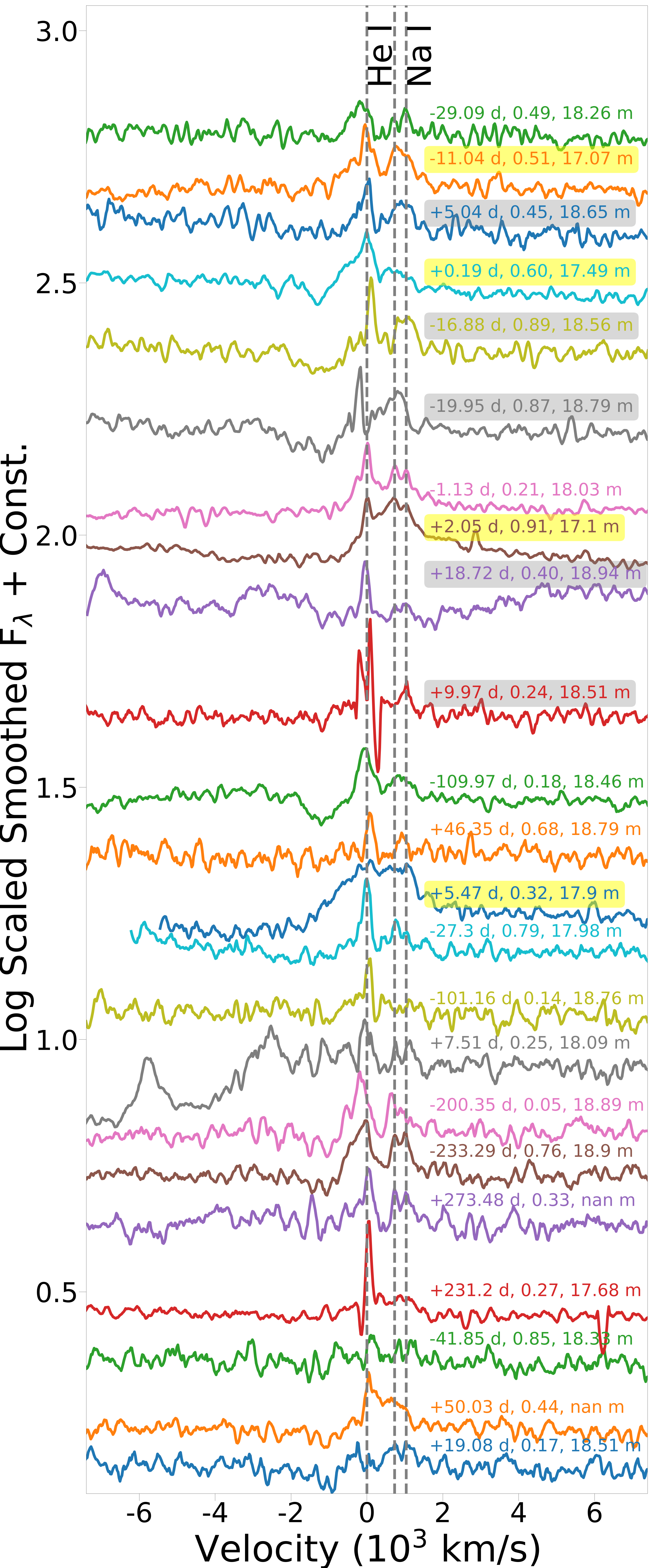}}
\end{tabular}
\caption{Spectral evolution of the He~{\sc i} line profile observed with the Binospec and BlueChannel instruments. The spectra are binned (see text). Note that the spectrum from 2016-02-16 has been cropped at the end due to block instrumental artifacts. Over time, the strength and shape of the He~{\sc i} line change, displaying variations such as P~Cygni profiles, and at times appearing very strong or nearly absent. The left panel shows narrow components on top of broad components due to CSM emission.}\label{fig:HeI}
\end{figure*}

\begin{figure*}
\begin{tabular}{cc}
\subfloat[BlueChannel 300] {\label{fig:BCH300HbetaNaIHeI}
    \includegraphics[height=0.87\linewidth]{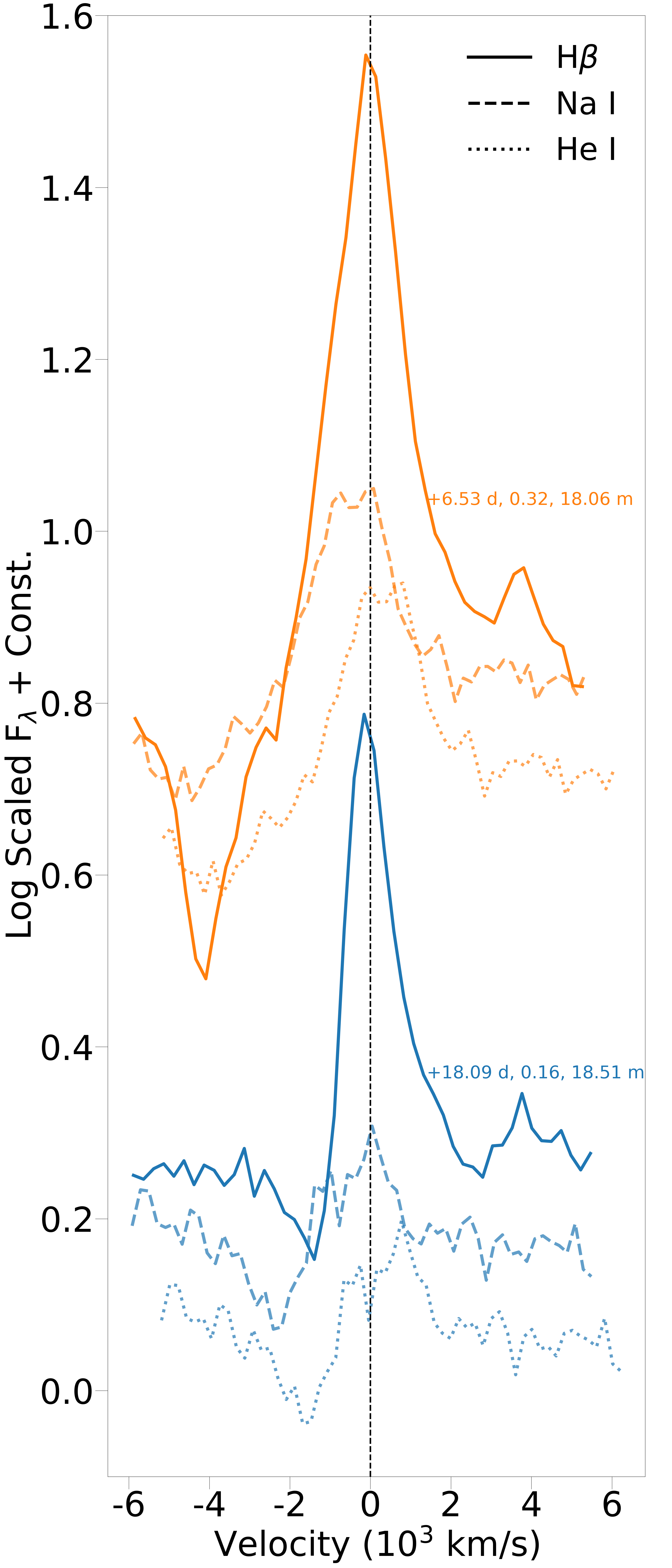}}&
    
    \subfloat[DEIMOS, Kast, and LRIS]{\label{fig:UCBHbetaNaIHeI}
    \includegraphics[height=0.87\linewidth]{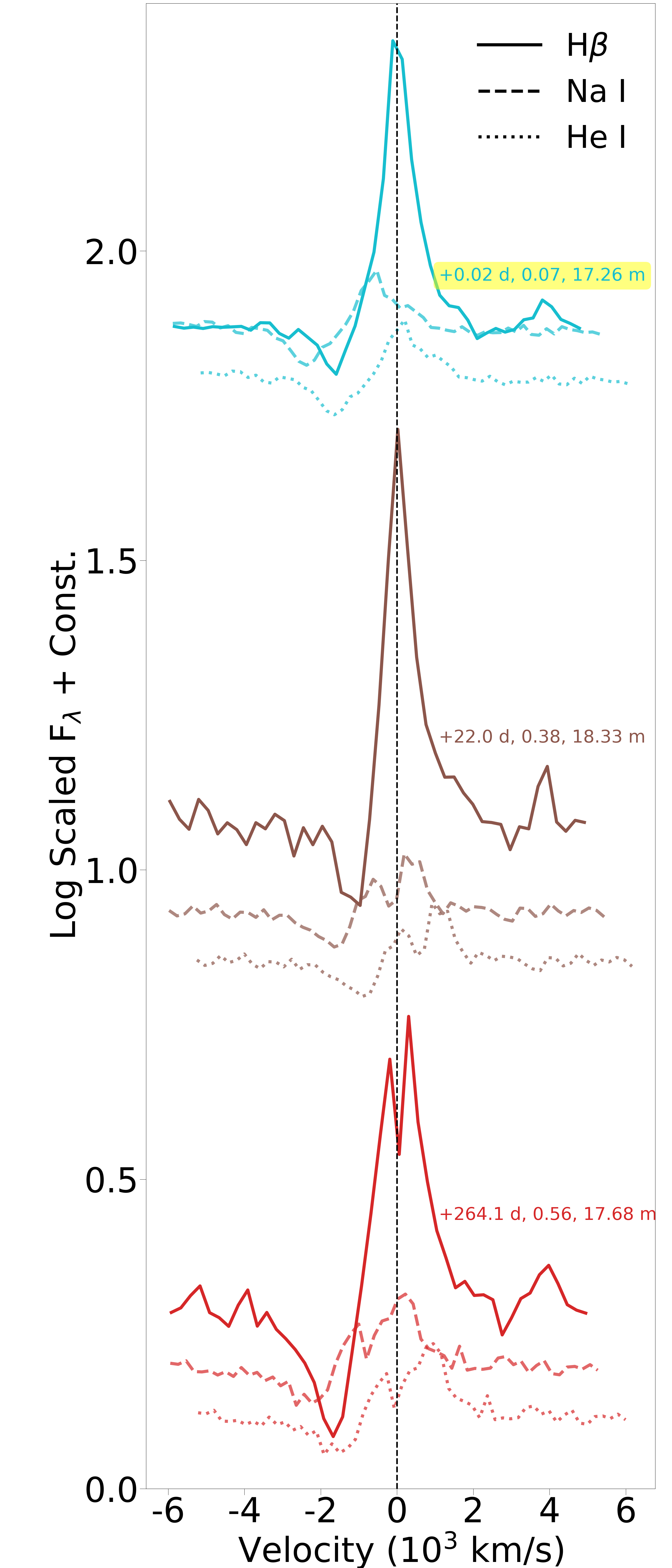}}
\end{tabular}
\caption{Velocity profiles of the H$\beta$ and He~{\sc i} + Na~{\sc i} lines. The solid line represents the H$\beta$ profile, while the He~{\sc i} + Na~{\sc i} blend is shown with non-solid lines: dotted lines indicate features centred at the He~{\sc i} rest frame, and dashed lines at the Na~{\sc i} rest frame. The P~Cygni absorption feature aligns more closely with H$\beta$ when centred at the He~{\sc i} rest velocity rather than at the Na~{\sc i} rest velocity.}\label{fig:HbetaNaIHeI}
\end{figure*}

\subsection{He~{\sc i} Lines}
Fig.~\ref{fig:HeI} shows various He~{\sc i} velocity profiles, with all spectra binned with a bin size of 3~\AA. Fig.~\ref{fig:BinoHeI} shows Binospec spectra, revealing strong He~{\sc i} features in both outburst and quiescent states, except on 2022-12-23 when the spectrum is significantly noisy. The high resolution of the instrument allows for the identification of both a narrow component around the He~{\sc i} rest velocity and a broad component. Since bright Na~{\sc i} emission is not expected in H~{\sc ii} regions, the narrow component in this case might be attributed to CSM emission. Additionally, Fig.~\ref{fig:BCH1200HeI} shows spectra from the BlueChannel 1200 grating. On dates such as 2019-06-04 and 2017-05-20, the spectra exhibit strong He~{\sc i} features during the outburst state, while on other dates like 2012-04-17, the He~{\sc i} feature appears very weak or nearly absent.

Fig.~\ref{fig:HbetaNaIHeI} shows the He~{\sc i} velocity profiles from BlueChannel 300 spectra, along with selected spectra from Lick and Keck data. The spectra are binned with a bin size of 4~\AA. The B\&C 300 He~{\sc i} velocity profiles are not shown owing to low resolution; however, Fig.~\ref{fig:bok300} indicates that weak He~{\sc i} features are also detected in B\&C 300 spectra. In Fig.~\ref{fig:HbetaNaIHeI}, H$\beta$ is shown as a solid line. The He~{\sc i} + Na~{\sc i} feature is centred using the He~{\sc i} rest wavelength (dotted line) and Na~{\sc i} rest wavelength at 5889.95~\AA\ (dashed line). The alignment of the He~{\sc i} + Na~{\sc i} feature appears consistent with the He~{\sc i} rest wavelength. Most likely, the absorption features are associated with He~{\sc i}, while the emission features are a combination of He~{\sc i} and Na~{\sc i}. Fig.~\ref{fig:BCH300HbetaNaIHeI}, and \ref{fig:UCBHbetaNaIHeI} also reveal that both the H$\beta$ and He~{\sc i} + Na~{\sc i} features display P~Cygni profiles simultaneously.

\begin{figure*}
\begin{tabular}{cc}
\subfloat[]{\label{fig:fwhmbch1200}
    \includegraphics[width=0.5\linewidth]{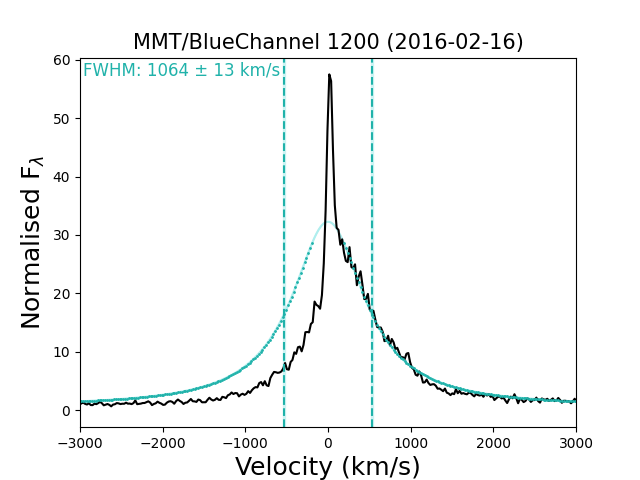}}&

     \subfloat[]{\label{fig:fwhmDeimos1200}
    \includegraphics[width=0.5\linewidth]{
    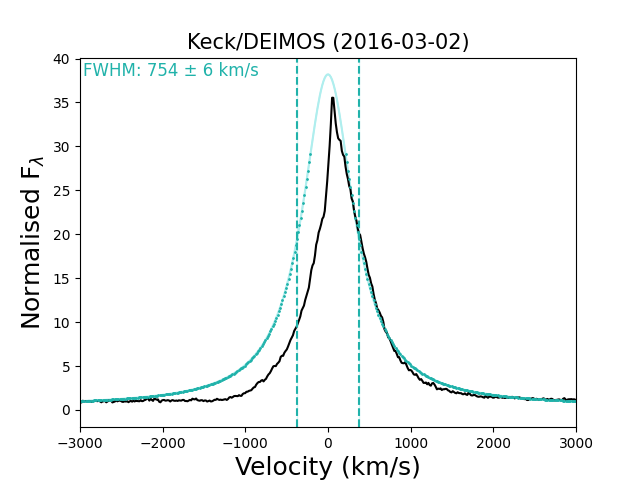}}\\
\subfloat[]{\label{fig:fwhmbok1200}
    \includegraphics[width=0.5\linewidth]{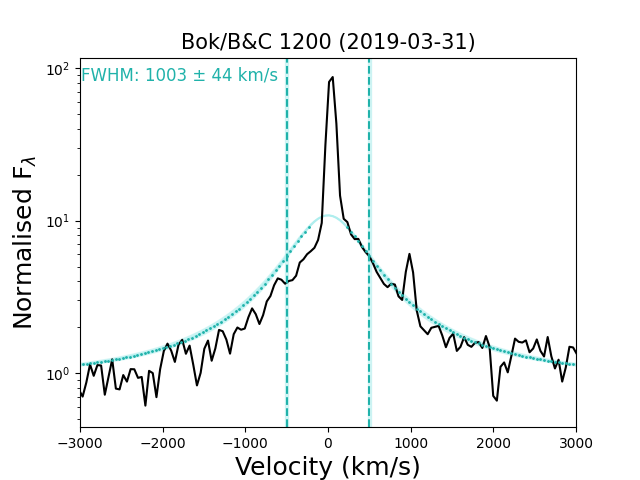}}&
        \subfloat[]{\label{fig:fwhmbino}
    \includegraphics[width=0.5\linewidth]{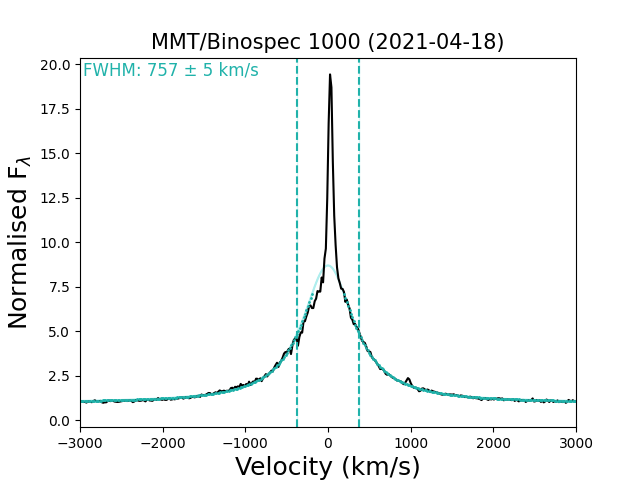}}\\

\end{tabular}
\caption{A few H$\alpha$ profiles are overlaid with dotted lines representing the combined Lorentzian and constant-background fits. Owing to contamination from the nearby H~{\sc ii} region or CSM, we omit the central peak section in our fitting process. Additionally, only the red wings (the redshifted wing reflected over the blue side) are fitted, while constraining the fit to be centred at zero. 
}\label{fig:fwhmselected}
\end{figure*}

\section{H\texorpdfstring{$\alpha$}{alpha} profile fit} \label{sec:fwhm}
Fig.~\ref{fig:fwhmselected} shows that the H$\alpha$ emission lines have a Lorentzian profile rather than a Gaussian profile. This broadening could be a result of electron scattering or the presence of moving material, or both, though not moving at velocities comparable to those of SN ejecta. To estimate the FWHM of the H$\alpha$ lines, we combine a Lorentzian function, representing the spectral line, with a constant background to account for the underlying continuum. Fig.~\ref{fig:fwhmselected} shows examples of Lorentzian fits applied to selected spectra.

To fit the broader emission features and minimise the influence of the narrow emission peak at the centre of the H$\alpha$ profile, we exclude the data points in close proximity to the H$\alpha$ narrow peak -- specifically, within $\pm$ 160~km s$^{-1}$ for the high-resolution spectra. The extent of the area excluded varies with the resolution of the instrument, which affects the estimated FWHM of H$\alpha$, making it either narrower or broader.
Given that some of the H$\alpha$ profiles are asymmetric with a red hump, we mirrored the red wing of the line profile over to the blue side to create a symmetric profile. This method was consistently applied to all spectra to maintain uniformity in our FWHM analysis. Consequently, the estimated FWHM can be larger if the red side is stronger (see Figs.~\ref{fig:fwhmbch1200}, \ref{fig:fwhmDeimos1200}, and ~\ref{fig:fwhmbok1200}) or smaller if the red side is weaker (see Figs.~\ref{fig:fwhmbino}). The Lorentzian fits, shown by dotted points in the figures, are confined to the velocity range of $-5000$ to 5000 km s$^{-1}$. 

We use the curve-fit function from the SciPy library in Python to perform the fits. The FWHM of the Lorentzian components was determined from the width parameter derived from these fits. Vertical dashed lines in the figures mark the FWHM of the Lorentzian peaks. The uncertainty in the FWHM was estimated by taking the square root of the diagonal elements of the covariance matrix associated with the fit width, with an additional term for residual variance to account for model-data deviations. The uncertainty in the FWHM is visually represented by shaded bars around these dashed lines, indicating potential variability due to fitting errors. Furthermore, uncertainties in the parameters of the Lorentzian fit, particularly the width, are illustrated by shaded regions around the main fit lines, which are estimated by adjusting the width parameter by $\pm$ the uncertainty of the width. It is important to note that FWHM uncertainties do not account for uncertainties due to instrument resolution, meaning that FWHM estimates for lower-resolution spectra are much more uncertain than those for high-resolution ones.

\begin{figure*}
\includegraphics[width=\textwidth]{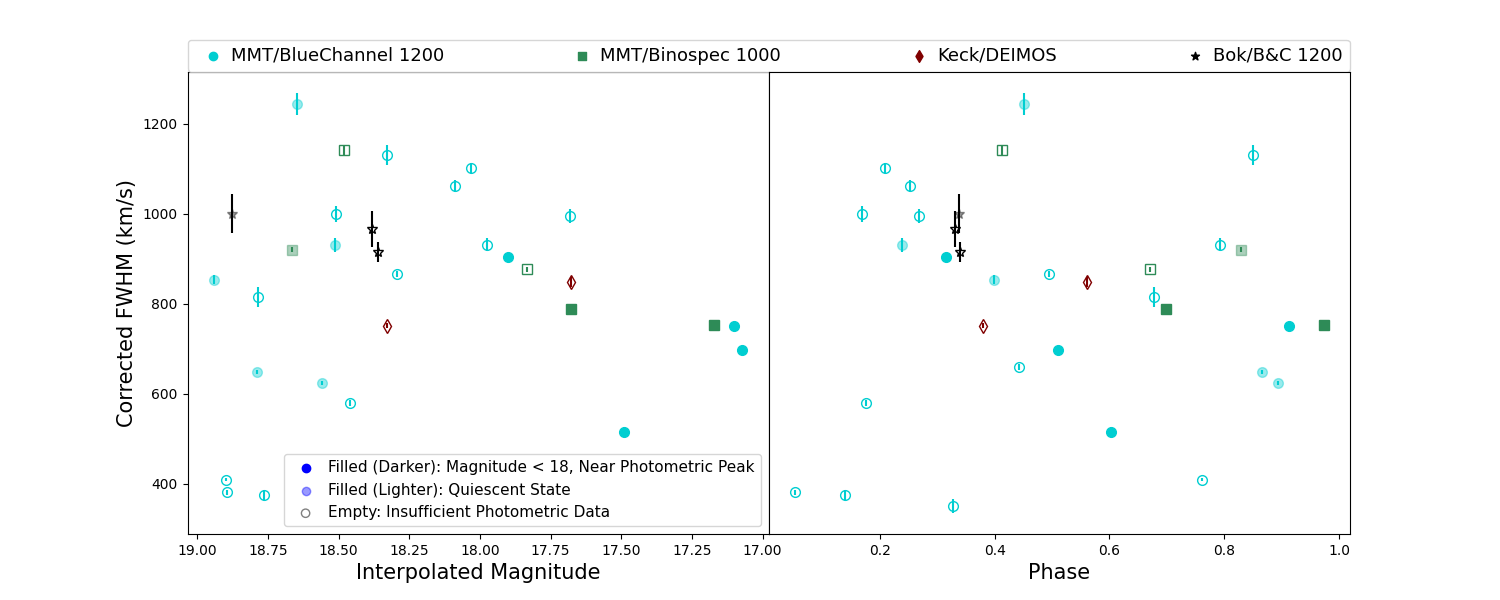}
\caption{The left and right panels show the FWHM of only high-resolution spectra, after correction for instrumental resolution, versus the interpolated magnitudes at the time the spectra were obtained, and the phase using the 113~d period, respectively. The corrected FWHM for the H$\alpha$ line ranges from $\sim$300 to 1300 km s$^{-1}$. Filled symbols with darker shades represent spectra with an interpolated magnitude brighter than 18 and observed close to the photometric peak, while filled symbols with lighter shades correspond to spectra taken during the quiescent state. Empty symbols represent spectra for which the system's state could not be determined owing to insufficient photometric data.  The shapes and colours indicate different instruments, as noted in the legend. No apparent correlation is found between the FWHM values and the eruptive or quiescent states, nor with the phase.}
\label{fig:Allfwhm}
\end{figure*}

Fig.~\ref{fig:Allfwhm} shows the estimated FWHM of only high-resolution spectra. The FWHM values have been corrected for instrumental resolution and are plotted against the interpolated magnitudes at the time the spectra were obtained (left panel), and the $\sim$113~d period phase cycle (right panel). No clear correlation is observed between the FWHM values and either the eruptive or quiescent states, or with the phase. Instrument resolutions are listed in Section~\ref{sec:obsspec}. Spectra with interpolated magnitudes brighter than 18, obtained within 15 days of a documented peak in the light curve, are plotted with filled dark shade symbols. Spectra taken during quiescent states are shown with lighter shade filled symbols. Corrected FWHM values vary between $\sim 300$ to 1300 km~s$^{-1}$. In the standard LBV scenario, one might generally expect slower outflows during brighter states because the star's radius increases, making it cooler and yielding lower escape velocities. While Fig.~\ref{fig:Allfwhm} does not show a strong trend, it seems that the upper threshold of velocities does decrease at brighter magnitudes.   Owing to gaps in the photometry data, we are unable to constrain the state of the system during most of the spectroscopic observations (shown with empty symbols). The measured FWHM is sometimes comparable to other LBVs, yet it can reach velocities similar to those observed in precursor of SN~2009ip and SN~2000ch. In the next section, we compare the spectra of AT~2016blu with those of SN~2000ch and the precursor of SN~2009ip.

\begin{figure*}
\subfloat[]{
  \includegraphics[width=0.8\textwidth]{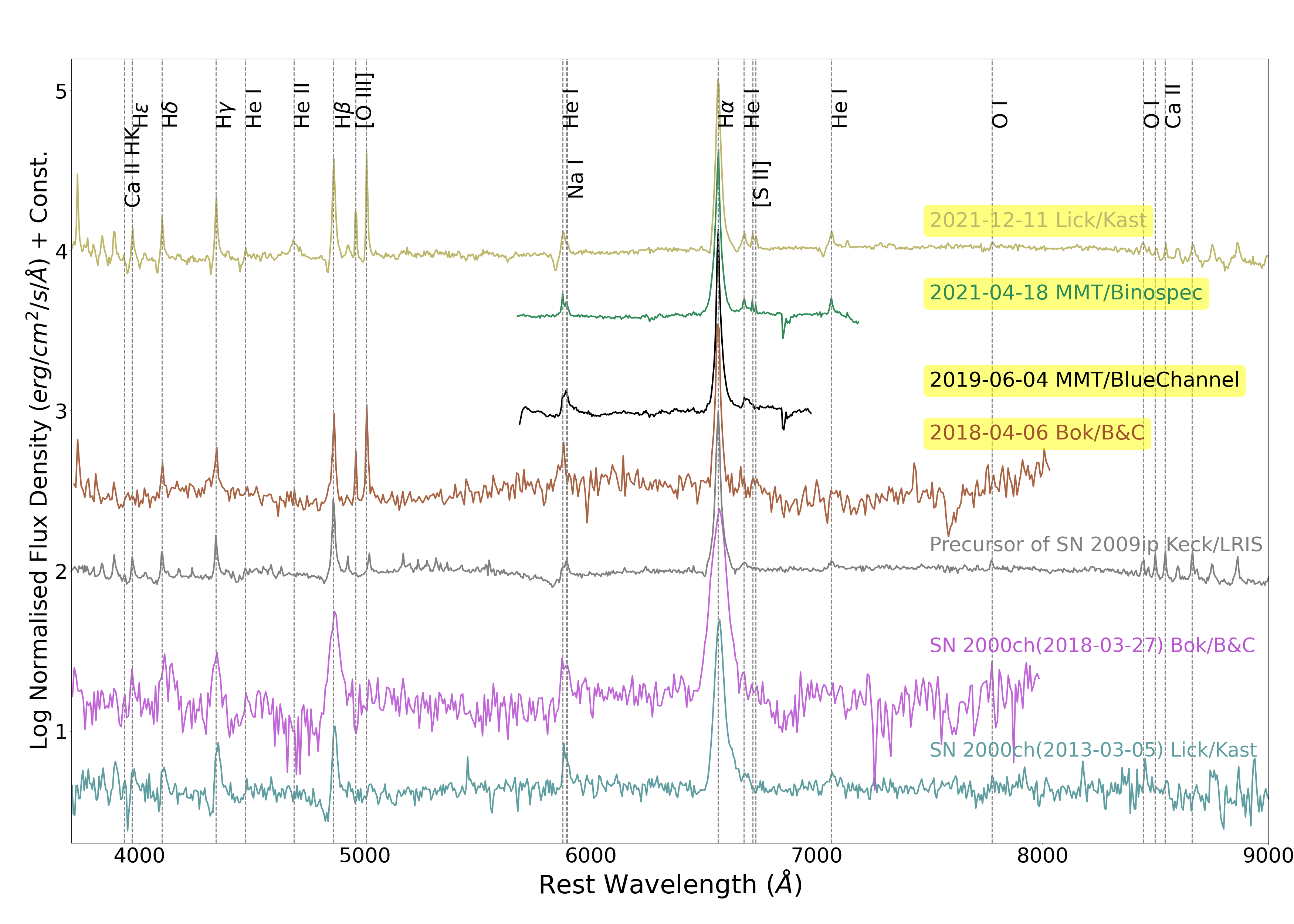}
  \label{fig:ComparisonOutburst}}\\
\subfloat[]{
  \includegraphics[width=0.8\textwidth]{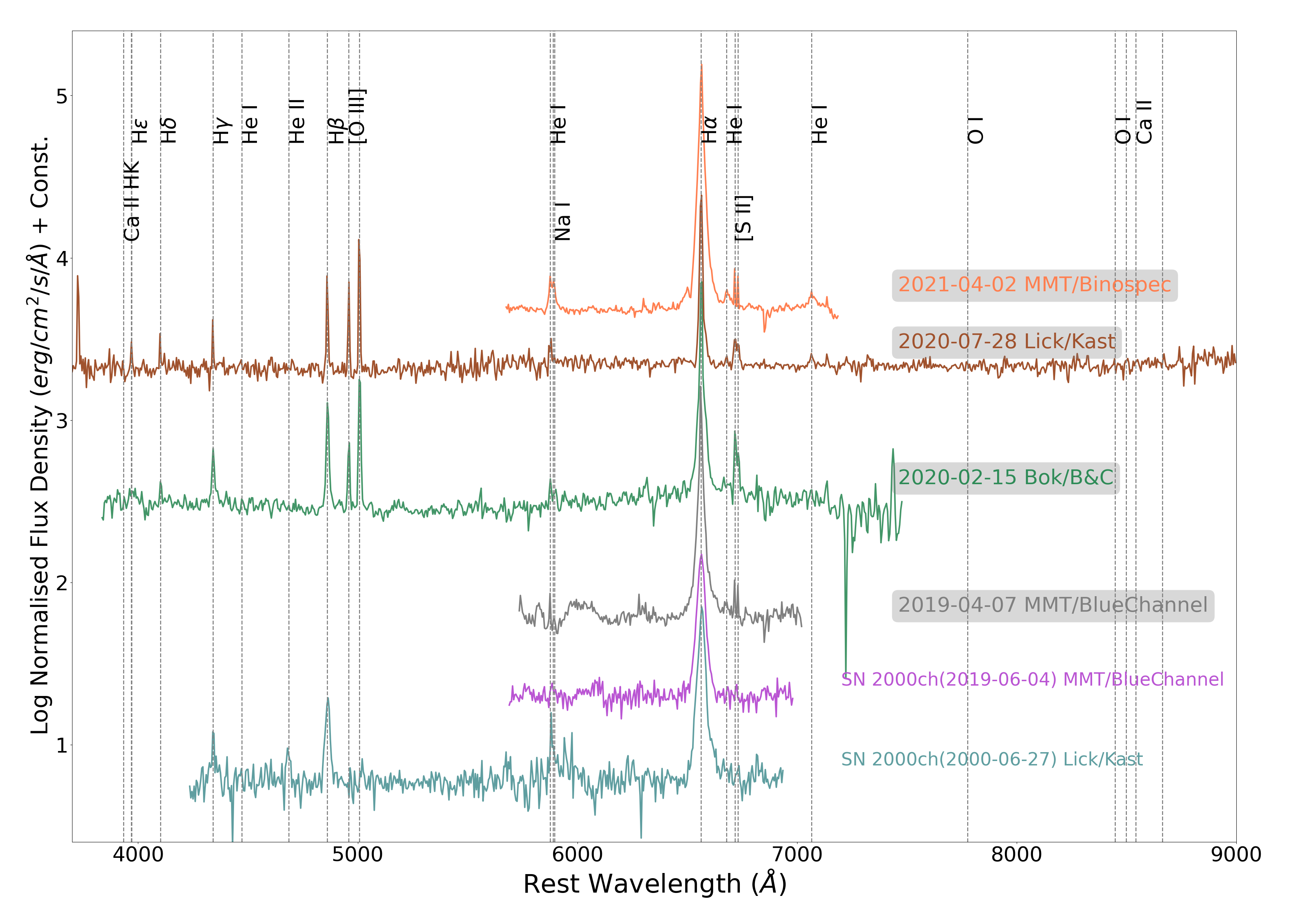}
  \label{fig:ComparisonQ}}
  
\caption{Comparison of normalised spectra from different telescopes during AT~2016blu's outburst and quiescent states, shown in the top and bottom panels, respectively. The spectra of AT~2016blu are compared with those of SN~2000ch, and the post-outburst spectrum of SN~2009ip. The spectrum of SN~2000ch from  2000 June is from \citet{W04}.
}\label{fig:comparison}
\end{figure*}

\section{Comparison with other SN impostors}\label{sec:comparison}
In Fig.~\ref{fig:comparison}, we compare the spectra of AT~2016blu with those of other known SN impostors. The spectra are binned using a bin size ranging from 5 to 8~\AA, selected based on the noise level of each individual spectrum. We have corrected the spectra of both SN~2000ch and the precursor of SN~2009ip for their respective redshift values, 0.0017 and 0.0059. The spectra of SN~2000ch and the post-outburst (but pre-SN) spectrum of SN~2009ip are also characterised by strong Balmer emission lines with broad wings reaching to high velocities. The spectrum of AT~2016blu includes [S {\sc ii}] and [O~{\sc iii}] emission lines, indicative of flux contamination from nearby H~{\sc ii} regions, which are absent in the spectra of the precursor of SN~2009ip and SN~2000ch. 

During the outburst state, the He~{\sc i} line at 5876~\AA\ is visible in all spectra; see Fig.~\ref{fig:ComparisonOutburst}. However, the He~{\sc i} line at 6678~\AA\ is detected in most spectra, but it appears weaker.  The Lick/Kast spectrum of AT~2016blu from December 11, 2021, displays Ca~{\sc ii} features similar to those observed in the precursor of SN~2009ip, which are absent in the spectra of SN~2000ch. Similarly, the He~{\sc ii} line at 4686~\AA\ is detected in the 2021 spectra but not in others. It is important to note that these lines are generally observed in the spectra of SN~2000ch \citep{W04,P10,S11}.
 
In the quiescent state, it is much more challenging to detect spectral lines because the targets are usually very faint.  For example, while the He~{\sc i} line at 5876~\AA\ is visible in all spectra, its presence in the SN~2000ch spectrum observed on 2019-06-04 is difficult to confirm owing to low signal-to-noise ratios. Fig.~\ref{fig:ComparisonQ} illustrates that although the higher Balmer lines exhibit a P~Cygni profile during an outburst, this feature is less pronounced during the quiescent state.

\begin{figure*}
\begin{tabular}{cc}
\subfloat[] {\label{fig:ComparisonOutburstHalpha}
\begin{overpic}[width=0.4\linewidth]{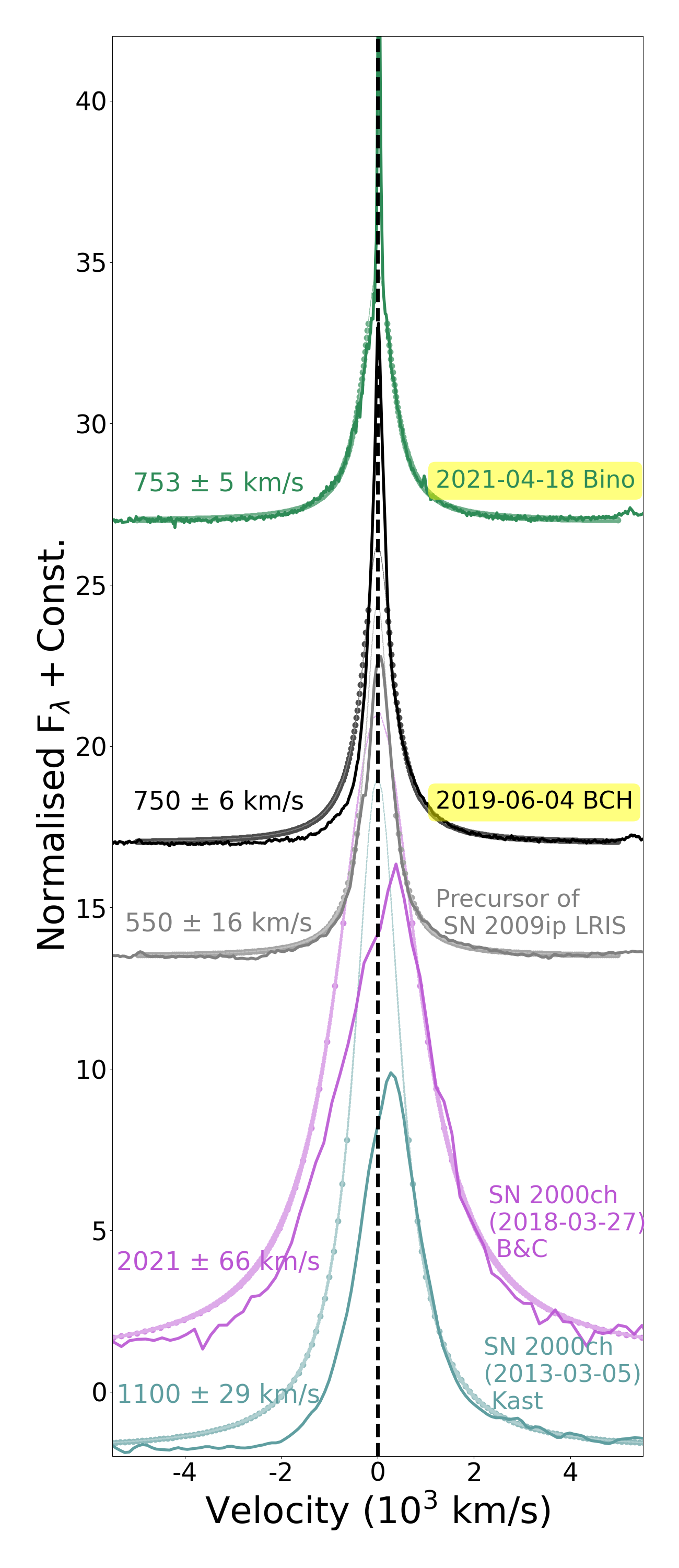}
            \put(9,96){\makebox(1,1){\textcolor{black}{\LARGE H$\alpha$}}}
        \end{overpic}}&
    \subfloat[]{\label{fig:ComparisonQHalpha}
\begin{overpic}[width=0.4\linewidth]{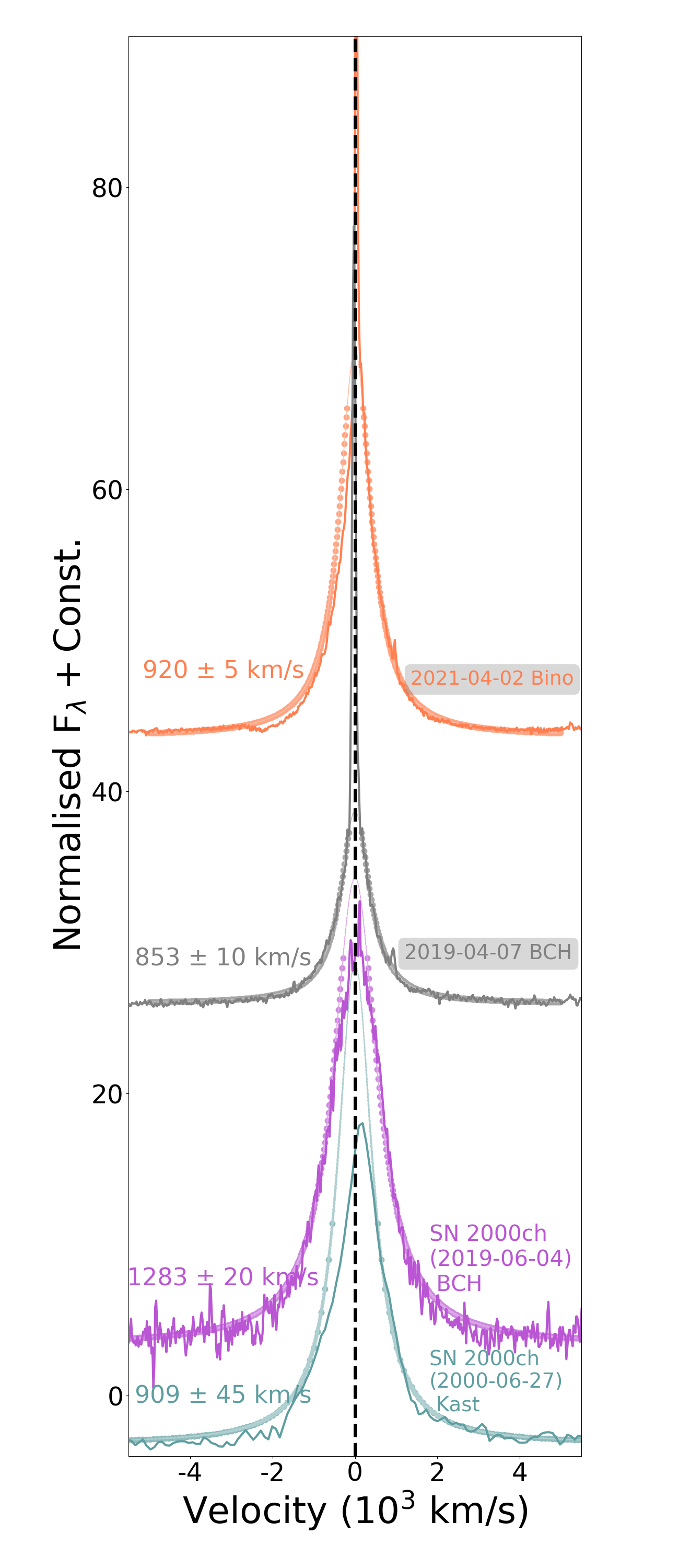}
            \put(10,96){\makebox(1,1){\textcolor{black}{\LARGE H$\alpha$}}}
        \end{overpic}}

\end{tabular}
\caption{H$\alpha$ velocity profile comparisons of only high-resolution spectra of AT~2016blu with SN~2000ch and the precursor of SN~2009ip. The H$\alpha$ peak of SN~2000ch is notably redshifted, which is also observed in some spectra of AT~2016blu when the narrow peak is excluded. Lorentzian fits (fitted the red wing only, excluding the central peak) and corrected FWHM estimates are provided for each spectrum. Typically, SN~2000ch exhibits a higher FWHM than AT~2016blu or the precursor of SN~2009ip. The FWHM of AT~2016blu varies over time, although these variations do not correlate with its eruptive state or 113~d phase. 
}\label{fig:ComparisonHalpha}
\end{figure*}

Fig.~\ref{fig:ComparisonHalpha} shows the velocity profiles for H$\alpha$ from only the high-resolution spectra shown in  Fig.~\ref{fig:ComparisonQ}. The H$\alpha$ peak in SN~2000ch appears redshifted relative to those of AT~2016blu and the precursor of SN~2009ip. However, if we disregard the narrow components of AT~2016blu's H$\alpha$ emission, the broad components of some H$\alpha$ lines sometimes also appear redshifted, similar to SN~2000ch (see Fig.~\ref{fig:halphahighres}). To maintain consistency with the analysis described in Section~\ref{sec:fwhm}, we exclude the narrow peak of the H$\alpha$ profile from all our fits and only fit the red side of the profile. The corrected Lorentzian FWHM velocities for the broad component of H$\alpha$ are labeled in the figures (see Fig.~\ref{fig:ComparisonOutburstHalpha}, and ~\ref{fig:ComparisonQHalpha}). For the SN~2000ch spectra, which extend to much larger velocities, we applied Lorentzian fits over an extended velocity range ($-8000$ to 8000 km s$^{-1}$) to accurately determine the FWHM. The FWHM for SN~2000ch is sometimes much larger than that for AT~2016blu. By contrast, the FWHM of the precursor of SN~2009ip is similar to that of AT~2016blu, regardless of whether the narrow H$\alpha$ peak is included or excluded. 

\section{Discussion}\label{sec:Discussion}
\citet{A16blu} proposed that AT~2016blu is an eccentric binary system where the primary star is an LBV, and the recurrent outbursts observed in AT 2016blu result from interactions between the two stars during periastron passages. The nature of the companion remains unclear, however. The specific type of binary interaction responsible for the brightening events depends on the companion's nature and may involve various physical phenomena, such as strong wind collisions, disc interactions, mass accretion onto a compact object, etc. The spectroscopic analysis of AT~2016blu provides several key insights into this system, beyond what has been inferred from the light curve alone.

(1) The spectrum of AT~2016blu is different from that of a typical LBV spectrum. The absence of [Fe~{\sc ii}] emission lines, which are often prominent in LBV spectra due to their dense, low-ionisation winds, suggests a different ionisation state or density.  AT~2016blu more closely resembles the hottest known LBVs, but it appears to stay hot even in outbursts.  In this sense, it is similar to a subset of LBVs like MCA-1b and Romano's star in M33, and HD~5980 in the SMC \citep{smith20}. Additionally, AT~2016blu shows H$\alpha$ FWHM values that are on the high end of the distribution for LBVs \citep{S11}, indicating a relatively fast outflow (or very dense wind; see below) compared to typical LBVs. Hotter stars generally tend to have faster winds than cooler stars, mostly because cooler stars are also larger and have lower escape speeds. However, other effects like accretion and shock excitation may also play a role here. The presence of these shock waves could result in a higher ionisation state of the gas and might explain the absence of [Fe~{\sc ii}] lines.

(2) The overall spectrum of AT~2016blu remains largely the same over time, with subtle changes. For example, occasional excess flux on the red side of the H$\alpha$ wing, variations in P~Cygni features, and the occasional presence of spectral lines like O~{\sc i}, Ca~{\sc ii}, and Fe~{\sc ii} point to variable conditions within the stellar environment. The P~Cygni profiles indicate strong stellar winds and outflows, while the presence of different ionisation states of elements suggests varying temperatures and densities. These observations are consistent with the characteristics of a star that undergoes episodic outbursts and mass ejections. 

(3) In AT~2016blu’s spectra, several epochs reveal an excess flux on the red wing of H$\alpha$, while a few also show an excess on the blue wing. This alternating asymmetry in the line profiles implies that either the ejected material is distributed in a nonuniform manner or the observed Doppler shifts result from orbital motion within the binary system.

\begin{figure*}
\includegraphics[width=0.7\textwidth]{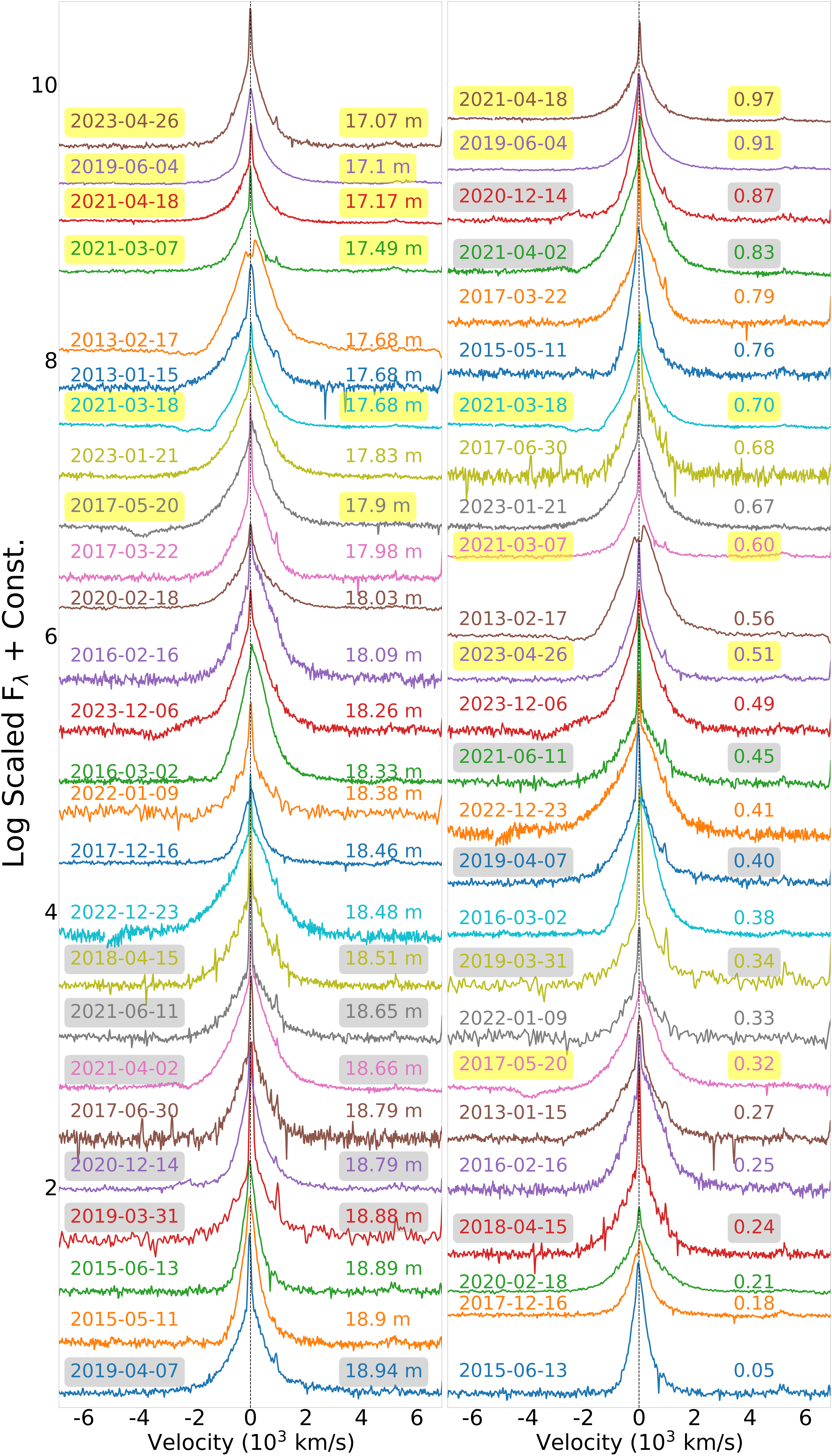}
\caption{The left and right panels show high-resolution H$\alpha$ profiles organised by magnitude and by 113~d phase, respectively. There is no observed correlation between either the high-velocity P~Cygni absorption or the red-side hump features with magnitude or phase. This may suggest an additional component beyond the binary nature of the system, such as the presence of ejected shells, or clumpy CSM.}\label{fig:halphasorted}
\end{figure*}

\begin{figure}
\includegraphics[width=\columnwidth]{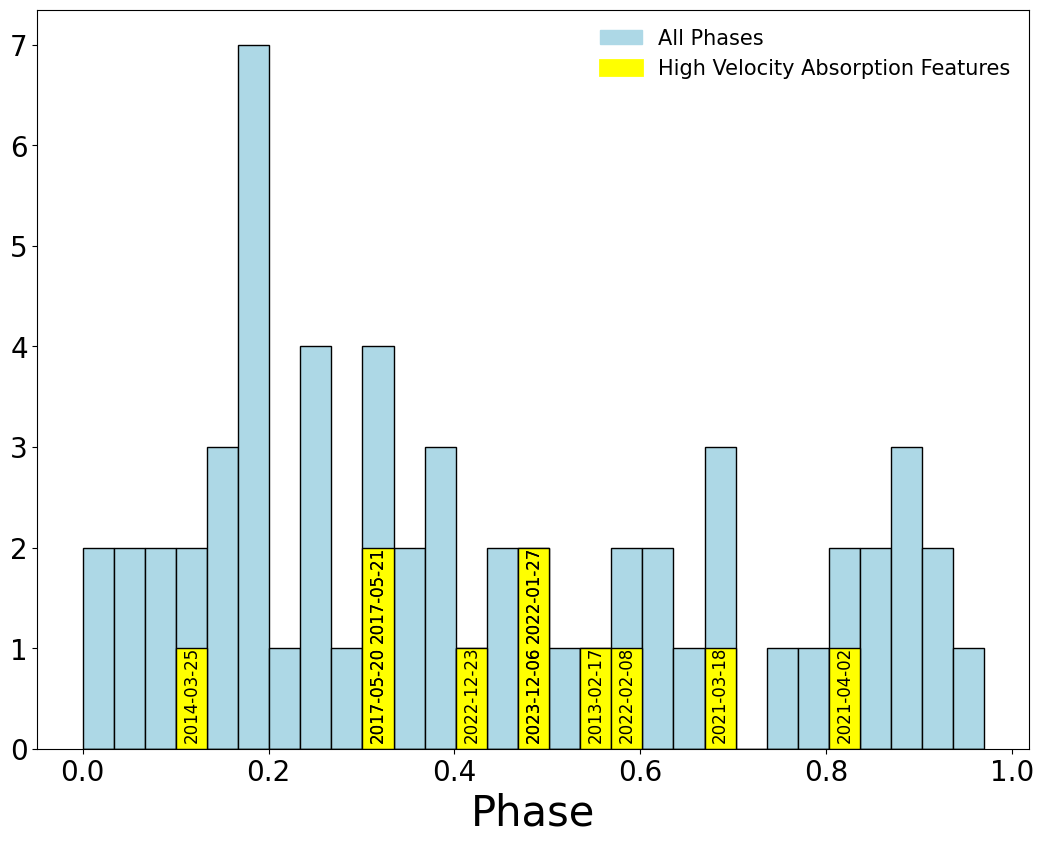}
\caption{Histogram of the phase using the 113~d period. Spectra with high-velocity absorption dips (greater than 2000~km~s$^{-1}$) are highlighted in yellow.  Absorption features are not correlated with the period.} 
\label{fig:HistPhase}
\end{figure}

(4) P~Cygni absorption features occasionally appear at very high velocities compared to FWHM estimates for emission components, suggesting that the emission and absorption components are not necessarily tracing the same gas in the outflow. This discrepancy may indicate the presence of a companion star with a faster outflow, which may be observed in absorption at certain phases. If the system had a purely spherical wind, the emission width should be similar to the absorption width along the line of sight. However, for instance, on 2017-05-20, the H$\alpha$ emission line has an FWHM of around 900 km~s$^{-1}$, while the absorption feature is observed at approximately $-$4000 km~s$^{-1}$, suggesting a different source or a shock from the fast outflow of the companion accelerating material. If due exclusively to a companion, one might expect these features to recur at similar phases in the orbital cycle. Fig.~\ref{fig:halphasorted} shows the high-resolution H$\alpha$ profiles sorted by magnitude and  113~d orbital phase. No clear correlation is observed between the high-velocity P~Cygni features and either the orbital phase or the magnitude. Fig.~\ref{fig:HistPhase} also shows a histogram of the orbital phase for all spectra, with spectra exhibiting absorption dips exceeding 2000 km~s$^{-1}$ highlighted in yellow. Once again, no correlation is observed, suggesting that these features may not be solely due to the influence of the companion.

(5) While the H$\alpha$  FWHM is commonly interpreted as a direct indicator of wind outflow speed, it is important to consider scenarios where electron scattering plays a dominant role, owing to the fact that the broad component of H$\alpha$ is approximated by a Lorentzian profile shape. If the H$\alpha$ line width is primarily due to electron scattering, the FWHM may not accurately represent the wind outflow speed (the actual outflow speed would be slower). Instead, the width may be more closely related to the thermal velocities of the electrons responsible for the scattering and the high optical depth of the material. In $\eta$ Carinae's current wind, the H$\alpha$ line exhibits a Doppler-broadened core with weaker electron-scattering wings at a mass-loss rate ($\dot{M}$) of $\sim 1 \times 10^{-3}$ M$_{\odot}$ yr$^{-1}$. Given that the Lorentzian profiles in AT~2016blu are persistent throughout its period, this suggests that its mass-loss rate could be even higher than 10$^{-3}$ M$_{\odot}$ yr$^{-1}$, consistent with the Lorentzian H$\alpha$ line shapes seen in models or early-time spectra of some SNe with strong early CSM interaction \citep{groh14,shivvers15}. The persistence of Lorentzian profiles may indicate either a strong steady-state wind from the LBV, or alternatively, high densities in a continuous accretion disc onto a companion star.

(6) Broad P~Cygni absorption features sometimes exhibit single troughs and other times double-dip profiles. The presence of an absorption P~Cygni feature in the Balmer lines that appears and disappears, and sometimes exhibits a double-dip feature, might be attributed to a variety of phenomena.  Changes in the strength and velocity of the stellar wind can cause the absorption feature to vary in intensity and even disappear at times, but the high velocity of these absorption features seems at odds with arising in the LBVs wind alone, since the outflow speed of the LBV wind should be comparable to or slower than the FWHM values of H$\alpha$ emission.   The double-dip feature might occur due to complex wind structures or multiple ejection events with different velocities. In a binary system, if the companion star has a disc, jet, or its own wind, and then this companion passes in front of the LBV, it might also change the observed absorption features. Alternatively, if during an eruption the star ejects mass and forms a dense high-speed shell, it may also be seen in absorption after the eruption.

(7) He~{\sc i} emission lines are detected most of the time, except when the target is faint and the light is dominated by the nearby star cluster rather than the transient. The He~{\sc i} line profile and strength vary significantly over time, often showing multiple peaks, which may result from blending with the Na~{\sc i} line. The presence of both He~{\sc i} + Na~{\sc i} emission lines indicates a wide range of ionisation levels, perhaps suggesting that accretion luminosity or shocks may add to the stellar photoionisation of the environment within the system. 

If the  He~{\sc i} lines were due to excitation by colliding wind shocks, one might expect that they should rise as the system approaches periastron and moves into a state of outburst, and should then diminish as the system moves toward apastron. However, we observe that the He~{\sc i} lines are visible at essentially all phases, except for epochs with low signal-to-noise ratio, as noted above. If the  He~{\sc i} lines are due instead to an accretion disc around the companion star, then radiative excitation from this accretion might be able to persist longer than the main phase of the visual-wavelength eruption, perhaps lasting for a few months. If the  He~{\sc i} line turns on around the time of the outburst, lasts for a while, fades away, and then reappears during the next outburst, it may indicate the influence of an accretion disc. Considering this scenario, we can estimate the accretion luminosity using $L = GM \dot{M}/2R$. For an accretion luminosity of $10^6~L_\odot$, which would be competitive with the star's luminosity, and assuming a companion mass $M = 1~M_\odot$ and radius $R = 1~R_\odot$, we find an accretion rate of $\sim 0.07~M_\odot$~yr$^{-1}$. This is quite high, suggesting that significant wind mass loss from the primary would be necessary to supply this material. However, the required mass accretion rate could be reduced if the companion is smaller, like a black hole, or significantly more massive.

(8) To better understand this system, we examine how the observed features correlate with the time of periastron. For instance, if the absorption features show no clear correlation with the orbital period, it might indicate that the star is randomly ejecting shells. On the other hand, if these features consistently appear after the main peak (when the magnitude is fainter), it could suggest that more mass is ejected during eruptions. As shown in Fig.~\ref{fig:halphasorted}, there is no clear evidence that the spectral changes are phase-dependent or correlate with the interpolated photometric magnitude. This lack of correlation suggests that additional components beyond the binary interaction may influence the system, such as the presence of ejected shells or clumpy CSM. Interactions between the companion and these dense regions of CSM or shells could give rise to clusters of brightening events near the predicted period that complicate the observed variability. Infrared observations with {\it JWST} would be valuable for detecting dust in the system. The presence of nested dust shells from repeating periastron encounters, similar to those observed in WR~140 (see Paper I for more details), may potentially explain the lack of correlation with the phase or interpolated photometric magnitude.

\begin{figure}
\includegraphics[width=\columnwidth]{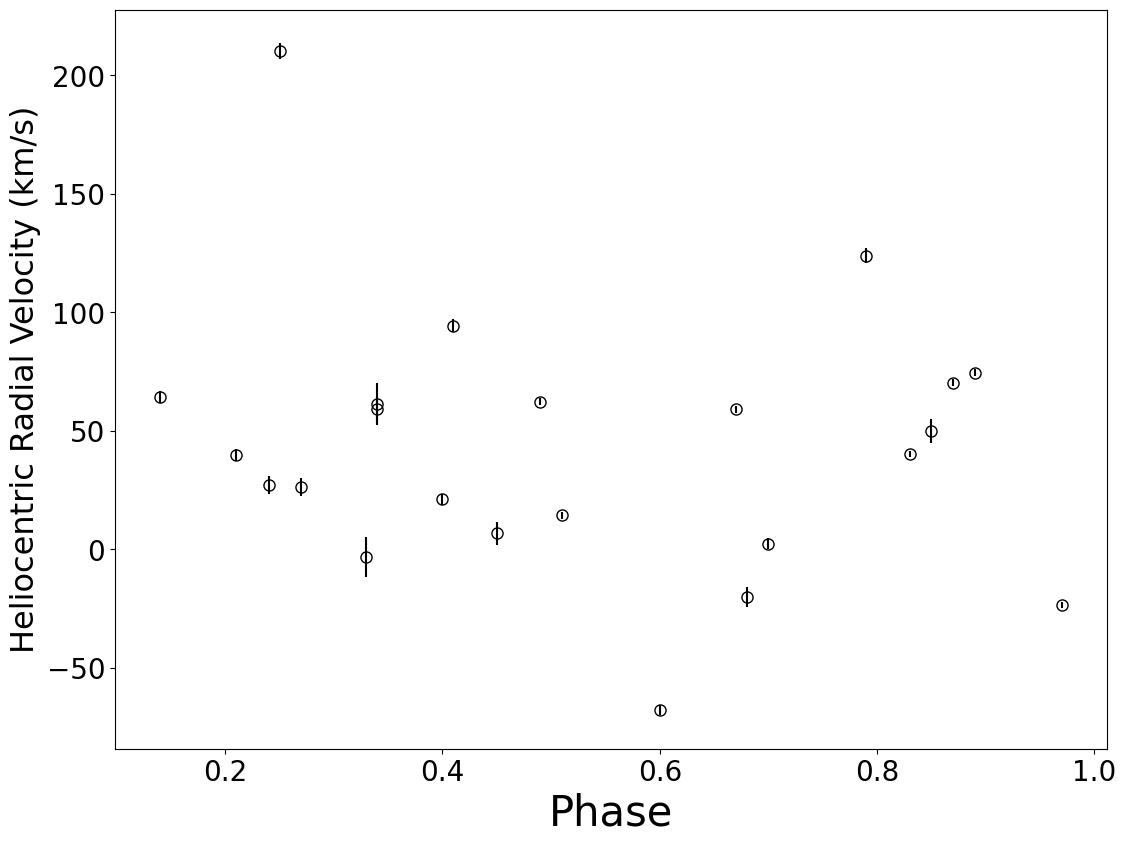}
\caption{Heliocentric radial velocities for the H$\alpha$ line, derived from high-resolution spectra obtained where the narrow component of the H$\alpha$ profile is resolved. The radial velocities exhibit significant variability but show no correlation with the orbital phase. The lack of a clear pattern suggests that other factors might be influencing the observed radial velocities, or that the variations may not be primarily due to binary motion.} 
\label{fig:RV}
\end{figure}

We also analyse the radial-velocity variations across the orbital phase. Since the system is likely an eccentric binary, these variations could be attributed to binary motion. Radial velocities were measured by fitting a Lorentzian profile to the broad component of the H$\alpha$ line and correcting for the heliocentric reference frame. To ensure accuracy, we restrict our analysis to high-resolution spectra, where the narrow H$\alpha$ component is well-resolved. This approach allows us to centre the line and minimise the impact of wavelength calibration errors on the observed velocity shifts. As shown in Fig.~\ref{fig:RV}, the heliocentric radial velocities display significant variability. However, this variability does not correlate with the orbital phase, suggesting that other factors might be influencing the observed radial velocities. Alternatively, as suggested in Paper I, AT~2016blu may be less eccentric than SN~2000ch, which could contribute to more erratic radial-velocity behaviour and obscure phase-dependent trends. 

Another possibility involves stochastic, repeating eruptions from a single star. However, this scenario also faces challenges in aligning with the observations. Such eruptions would likely cause the photosphere to expand and cool or lead to increased mass loss, neither of which have been observed. Furthermore, there is currently no model to explain how this process works or how the star becomes so luminous in such a short time without changing its spectrum. For instance, \cite{G21} proposed that quasiperiodic variability in LBVs can result from a wind-envelope interaction. However, this model involves temperature variations as the stellar envelope expands and contracts, which is not observed in AT~2016blu. Additionally, it predicts S~Dor variability on timescales of years to decades, far longer than AT~2016blu’s 113-day period, and far longer than the actual duration of bright outbursts of AT~2016blu within that 133~d cycle. These discrepancies suggest that wind-envelope interactions are unlikely to be the primary driver of its periodic outbursts.

\section{Conclusion}\label{sec:conclusion}
We present spectroscopic analysis of the SN impostor AT~2016blu, located in an outer spiral arm of NGC~4559. AT~2016blu experienced at least 19 outbursts in 2012–2022, recurring with a period of $\sim 113 \pm 2$~d \citep{A16blu}. Recent photometric data indicate that AT~2016blu experienced its \nth{21} outburst around 2023 May/June, as previously predicted based on its estimated period. These outbursts are suggested to be driven by interactions between two stars near periastron in an eccentric binary system, where the primary star is a massive LBV, and the companion could either be a compact object, such as a neutron star or black hole, or another star.

Our spectroscopic observations, covering multiple outbursts and quiescent states over the past 12~yr, document this transient's complex behaviour. In general, the spectra of AT~2016blu differ from those of a typical LBV spectrum, instead closely resembling the subset of LBVs that remain hot during eruptions \citep{smith20}, including MCA-1B and Romano's star in M33 and HD~5980 in the SMC.  The fact that these remain at a roughly constant temperature as they brighten requires a significant increase in bolometric luminosity, instead of a visual brightening due to redistribution of ultraviolet flux to shorter wavelengths, as usually assumed for LBVs. The AT~2016blu spectra show broad H$\alpha$ profiles with Lorentzian shapes and FWHM values that vary significantly, sometimes much higher than the typical FWHM of LBVs. The higher FWHM suggests the presence of strong stellar winds.  Occasionally, broad H$\alpha$ components with asymmetric profiles, featuring a red-side hump, are also observed, indicating self-absorption effects and changes in the density and velocity structure of the stellar wind. Overall, the appearance and behaviour of AT~2016blu's spectrum closely resemble the spectra of SN impostors like the precursor outbursts of SN~2009ip \citep{S10} before its 2012 core-collapse event, and also spectra of the SN impostor SN~2000ch \citep{W04,P10,S11}, which has very similar recurring periodic outbursts \citep{A00ch}.

P~Cygni absorption features are evident during both outburst and quiescent states, indicating the presence of strong stellar winds and episodic mass ejections. These features appear as either single or double P~Cygni profiles, reaching velocities significantly higher than the FWHM of the emission component of  H$\alpha$. This suggests that the absorption component originates from a different source, potentially due to strong winds from the companion star, the presence of dense shells, a jet, or an accretion disc seen along the line of sight to the LBV's wind.

The He~{\sc i} lines (where He~{\sc i} $\lambda$5876 is frequently blended with Na~{\sc i}) are detected in most spectra, indicating the coexistence of hot and cold gas within the system. The persistence of He~{\sc i} throughout most of the orbital period may suggest that excitation of the He~{\sc i} lines is driven by accretion onto a companion object rather than colliding wind shocks. X-ray observations are needed to determine whether the companion is a compact object or another massive star by detecting potential accretion-driven X-ray emission or colliding winds, providing strong evidence of an interacting binary system.

In conclusion, our study reveals that AT~2016blu is characterised by a complex and dynamic stellar environment, marked by significant spectral variations, strong winds, and interactions within an eccentric binary system. Continued high-cadence photometric and spectroscopic observations over a complete outburst cycle, from the quiescent phase through the peak and back to the end of the cycle, are crucial.  Such data would allow for a detailed comparison of the spectral and photometric evolution at each stage of variability, providing deeper insights into the underlying mechanisms driving the recurrent outbursts. 

Additionally, identifying more objects like AT~2016blu is critical for understanding the broader population of SN impostors and LBVs of this type. Currently, AT~2016blu and SN~2000ch are the only known examples exhibiting short quasiperiodic eruptions (unless we also count the now dead example of SN~2009ip). The upcoming Vera C. Rubin Observatory’s Legacy Survey of Space and Time (LSST) will provide an unprecedented opportunity to discover and characterise similar systems. Expanding this sample will help to demonstrate the range of orbital periods that experience this phenomenon, and will also help clarify the strictness of the periodicity in such systems (which may be related to orbital eccentricity or other factors, as discussed above).

\section*{Acknowledgements}
This research has made use of the NASA/IPAC Infrared Science Archive, which is funded by the National Aeronautics and Space Administration (NASA) and operated by the California Institute of Technology. It also used data from the Asteroid Terrestrial-impact Last Alert System (ATLAS) project, which is funded primarily to search for near-Earth objects (NEOs) through NASA grants NN12AR55G, 80NSSC18K0284, and 80NSSC18K1575; byproducts of the NEO search include images and catalogues from the survey area. The ATLAS science products have been made possible through the contributions of the University of Hawaii Institute for Astronomy, the Queen’s University Belfast, the Space Telescope Science Institute (STScI), the South African Astronomical Observatory, and The Millennium Institute of Astrophysics (MAS), Chile. We acknowledge ESA {\it Gaia}, DPAC, and the Photometric Science Alerts Team (http://gsaweb.ast.cam.ac.uk/alerts). 

Observations using Steward Observatory facilities were obtained as part of the large observing program AZTEC: Arizona Transient Exploration and Characterization. Some of the observations reported in this paper were obtained at the MMT Observatory, a joint facility of the University of Arizona and the Smithsonian Institution.

Some of the data presented herein were obtained at the W. M. Keck Observatory, which is operated as a scientific partnership among the California Institute of Technology, the University of California, and NASA; the observatory was made possible by the generous financial support of the W. M. Keck Foundation. The authors wish to recognise and acknowledge the very significant cultural role and reverence that the summit of Maunakea has always had within the indigenous Hawaiian community. We are most fortunate to have the opportunity to conduct observations from this mountain. 
A major upgrade of the Kast spectrograph on the Shane 3 m telescope at Lick Observatory, led by Brad Holden, was made possible through gifts from the Heising-Simons Foundation, William and Marina Kast, and the University of California Observatories. Research at Lick Observatory is partially supported by a generous gift from Google. 
We thank the staffs at the MMT, Lick, and Keck Observatories for their excellent assistance.
The following people helped with the Lick and/or Keck observations or reductions, for which we are grateful: Brad Cenko,
Ori Fox, Melissa Graham, Pat Kelly, Adam Miller, Jeffrey Silverman, and WeiKang Zheng.

Support for M.A. was provided by the VITA-Origins Fellowship, including funding from the Virginia Institute for Theoretical Astrophysics (VITA), supported by the College and Graduate School of Arts and Sciences at the University of Virginia.
Time-domain research by D.J.S. is supported by National Science Foundation (NSF) grants 2108032, 2308181, 2407566, and 2432036 and the Heising-Simons Foundation under grant \#2020-1864.
A.V.F.'s group at UC Berkeley has received financial assistance from the Christopher R. Redlich Fund, 
% Gary and Cynthia Bengier, 
%(S.B.C. was a Bengier Postdoctoral Scholar),  
William Draper, Timothy and Melissa Draper, 
Briggs and Kathleen Wood, Sanford Robertson (T.G.B. is a Draper-Wood-Robertson Specialist in Astronomy), Shawn and Oliva Atkisson, Sue Broadston and Jim Ostendorf, Jim Connelly and Anne Mackenzie, Curt and Shelley Covey, Heidi Gerster, Harvey Glasser, John Gnuse and Stacey Dunn-Emke, Jeff and Alison Holland,  Michael Kast and Rebecca Leon, Ken and Gloria Levy, Walter and Karen Loewenstern, Mike McCaw and Janet Westin, Cat Rondeau, Richard and Betsey Sesler, David and Joanne Turner, Gerry and Virgnia Weiss, and numerous other donors.

\section*{Data Availability}
ZTF data are available in the public domain \url{https://irsa.ipac.caltech.edu/Missions/ztf.html}. ATLAS data are available in the ATLAS Forced Photometry server \url{https://fallingstar-data.com/forcedphot/}. {\it Gaia} data are available in \url{http://gsaweb.ast.cam.ac.uk/alerts/home}.  The spectroscopic data underlying this article will be shared on reasonable request to the corresponding author.

\bibliographystyle{mnras}
\DeclareRobustCommand{\VAN}[3]{#3}
\bibliography{MA}
%\begin{thebibliography}{99}
%\end{thebibliography}

%%%%%%%%%%%%%%%%%%%%%%%%%%%%%%%%%%%%%%%%%%%%%%%%%%
%%%%%%%%%%%%%%%%% APPENDICES %%%%%%%%%%%%%%%%%%%%%
%\appendix

%%%%%%%%%%%%%%%%%%%%%%%%%%%%%%%%%%%%%%%%%%%%%%%%%%
% Don't change these lines
\bsp	% typesetting comment
\label{lastpage}
\end{document}